\documentclass[aps, prl, reprint, doublecolumn, nofootinbib, superscriptaddress]{revtex4-2}
\usepackage{amsmath}
\usepackage{amssymb}
\usepackage{amsfonts}
\usepackage{amsthm}
\usepackage{mathrsfs}
\usepackage{booktabs, array}
\usepackage{soul}
\usepackage[T1]{fontenc}
\usepackage[caption=false]{subfig}
\usepackage{graphicx} 
\usepackage{multirow}
\usepackage{makecell}
\usepackage[dvipsnames]{xcolor}
\usepackage{tikz}
\usetikzlibrary{arrows.meta,positioning,calc}

\newcommand{\DD}{\boldsymbol{\Delta}}
\newcommand{\bd}{\bar\delta}
\newcommand{\cS}{\mathcal S}

\definecolor{linkcolor}{rgb}{0.0,0.3,0.5}
\usepackage[
    hypertexnames=false,
    unicode,
    colorlinks=true,
    linkcolor=linkcolor,
    citecolor=linkcolor,
    filecolor=linkcolor,
    urlcolor=linkcolor,
    pdfusetitle
]{hyperref}
\usepackage{url}
\hypersetup{breaklinks=true}

\usepackage{orcidlink}

\newcommand{\UIUC}{Illinois  Center  for  Advanced  Studies  of  the  Universe \& Department of Physics, 
\\University of Illinois Urbana-Champaign, Urbana, Illinois 61801, USA}

\allowdisplaybreaks
\begin{document}
    \title{Metric Reconstruction for Generic Black-Hole Perturbations}
    \author{Dongjun Li}
    \email{dongjun@illinois.edu} 
    \affiliation{\UIUC}

    \author{Nicol\'{a}s Yunes}
    \affiliation{\UIUC}
    
    \date{\today}
    \begin{abstract}
        Most sourced radiation-gauge reconstructions of Kerr perturbations use Hertz potentials. We establish the first self-contained Newman--Penrose transport hierarchy for reconstructing metric perturbations with generic conserved sources on Ricci-flat Petrov type D backgrounds. In a traceful radiation gauge, two first-order equations determine the trace, after which the total metric follows from an extreme Weyl scalar and the prescribed Ricci tensor without a Hertz potential. We illustrate the method for a Schwarzschild black hole with a thin static shell, including a source-supported static completion sector.

    \end{abstract}
    \maketitle

\noindent\textit{Introduction---}Gravitational-wave observations have opened a direct window into gravity in the dynamical and nonlinear regime through compact-object mergers. With the improving sensitivity of current detectors \cite{LIGOScientific:2025slb} and the expected launch of future detectors \cite{Evans:2021gyd, Punturo:2010zz, LISA:2024hlh}, these observations will allow us to stress-test general relativity, search for new fundamental fields, and probe astrophysical environments around black holes. In two especially important regimes, the ringdown (i.e.~when the remnant of a compact binary collision settles down to a stationary state through gravitational-wave emission) and extreme mass-ratio inspirals (i.e.~when a small compact object spirals into a supermassive black hole), the dynamics can be accurately described as perturbations of a single black hole. Black hole perturbation theory therefore provides a controlled framework for isolating leading-order effects from different physical mechanisms, while systematically incorporating higher-order effects when needed.

Modern black hole perturbation theory originates from studying linear metric perturbations of Schwarzschild black holes, where each parity sector is governed by a decoupled differential equation \cite{Regge:1957td, Zerilli:1971wd, Moncrief:1974am}. However, this metric-based formulation does not generalize directly to rotating black holes due to the loss of spherical symmetry, motivating the curvature-based approach developed by Teukolsky \cite{Teukolsky:1973ha}. Focusing on two Weyl scalars, $\Psi_0$ and $\Psi_4$, which describe gravitational waves toward the horizon and null infinity, Teukolsky derived a decoupled and separable equation for each scalar. Remarkably, the radiative part of the metric perturbation can be reconstructed from either scalar \cite{Chrzanowski:1975wv, Cohen_Kegeles_1975, Wald:1978vm}, while the remaining completion pieces correspond to mass, angular-momentum, and gauge sectors. To date, the Teukolsky formalism underpins precision studies of ringdown \cite{Leaver:1985ax} and extreme mass-ratio inspirals \cite{Hughes:1999bq, Hinderer:2008dm} in vacuum general relativity, and has recently been extended to incorporate beyond-Einstein and matter effects \cite{Li:2022pcy, Hussain:2022ins, Cano:2023tmv}.

The ability to perform metric reconstruction not only reflects a fundamental property of vacuum general relativity but also has practical importance. When modeling extreme mass-ratio inspirals, metric reconstruction is essential for computing second-order self-force effects required for accurate waveform modeling over the $\mathcal{O}(10^4\text{–}10^5)$ cycles accessible to space-based detectors \cite{Campanelli:1998jv, Wardell:2021fyy, LISAConsortiumWaveformWorkingGroup:2023arg} and extending perturbative predictions toward the comparable-mass and merger regimes for ground-based detectors \cite{Wardell:2021fyy, Kuchler:2025hwx}. More broadly, metric reconstruction underlies extensions of the standard modeling approach for extreme mass-ratio inspirals to non-vacuum environments and beyond-Einstein theories \cite{Dyson:2025dlj, Li:2025ffh, Li:2026abc}, particularly within the modified Teukolsky formalism \cite{Li:2022pcy, LaHaye:2025ley, Li:2025ffh, Li:2026abc}. For ringdown, metric reconstruction enables the modeling of quadratic quasinormal modes \cite{Ma:2024qcv, Khera:2024bjs} and the computation of quasinormal-mode spectral shifts in modified gravity \cite{Cano:2024ezp, Aly:2026otj, Li:2025fci}.

The most widely used metric reconstruction scheme in vacuum general relativity is the Chrzanowski–Cohen–Kegeles (CCK) approach \cite{Chrzanowski:1975wv, Cohen_Kegeles_1975}, which expresses the metric perturbation $h_{\mu\nu}$ in terms of a Hertz potential $\Phi$ in tracefree radiation gauges, 
\begin{equation} \label{eq:radiation_gauges_tracefree}
    h_{\mu\nu}e_{a}^{\mu}=0\,,
    \quad h\equiv h_{\mu\nu}g^{\mu\nu}=0\,,
\end{equation}
where $e_{a}^{\mu}$ is a tetrad vector along either the ingoing ($n^{\mu}$) or outgoing ($l^{\mu}$) principal null direction of the black hole background. The Hertz potential $\Phi$ is related to either the Weyl scalar $\Psi_{0}$ or $\Psi_{4}$ through a fourth-order differential relation, with explicit inversion procedures developed by Lousto and Whiting for a Schwarzschild background \cite{Lousto:2002em}, and extended by Ori to a Kerr background \cite{Ori:2002uv}. 

Since the tracefree condition (i.e., $h=0$) is compatible only with vacuum or restricted sources, several source-inclusive extensions have been developed. One class introduces a corrector tensor so that the residual source admits a tracefree radiation-gauge Hertz reconstruction \cite{Green:2019nam, Toomani:2021jlo, Bourg:2024vre, Hollands:2024iqp}. Another approach patches vacuum CCK solutions on either side of a point-particle source and then transforms the resulting metric to Lorenz gauge \cite{Dolan:2021ijg, Dolan:2023enf}. Other curvature-based frameworks have also been constructed. In a Schwarzschild background, Lousto generalized the inverse Chandrasekhar relations to include matter and reconstructed the metric in Regge--Wheeler gauge for a generic source \cite{Lousto:2005xu}. Using the identities of Aksteiner, Andersson, and B\"ackdahl~\cite{Aksteiner:2016pjt}, Ref.~\cite{Nasipak:2025tby} implemented sourced radiation-gauge reconstruction for point particles on eccentric and precessing orbits around a rotating primary, while Ref.~\cite{Wardell:2024yoi} extended this approach to generic sources in Lorenz gauge. In particular, Ref.~\cite{Wardell:2024yoi} reconstructs generic-source Kerr perturbations from six sourced scalars up to a time integral, so the static sector requires separate treatment. Likewise, when the physical Weyl scalar is the input to a CCK-like reconstruction, the curvature-to-Hertz relation is algebraic for separated non-static modes in vacuum \cite{vandeMeent:2017bcc}, but source regions require solving the fourth-order relation with homogeneous and matching data, and the static sector is handled separately \cite{Toomani:2021jlo, Hollands:2024iqp}. Thus, CCK-like approaches retain an additional curvature-to-Hertz conversion step and organize static and completion sectors separately from the radiative Hertz potential.

In this work, we establish a different reconstruction architecture. Building on Chandrasekhar's foundational formulation \cite{Chandrasekhar_1983} and the later vacuum extensions in \cite{Loutrel:2020wbw, Ripley:2020xby}, we show for the first time that the Newman-Penrose (NP) commutation relations, Ricci identities, and Bianchi identities can themselves be organized into a self-contained hierarchy for a generic conserved source. A similar procedure was developed by \cite{Suvorov:2019qow} in $f(R)$ gravity specifically. In a traceful radiation gauge, two first-order transport equations determine the trace $h$, after which the total metric is reconstructed hierarchically from the physical $\Psi_{0,4}^{(1)}$ and the prescribed Ricci tensor through first-order transport equations, without introducing a Hertz potential. This direct physical-curvature interface, thus, treats radiative, static, and completion sectors within the same hierarchy.

The resulting framework provides a systematic tool for studying nonlinear perturbations in ringdown and extreme mass-ratio inspirals in vacuum general relativity, non-vacuum environments, and beyond-Einstein theories. The framework may be particularly advantageous for time-domain, zero- or low-frequency, and iterative effective-source calculations. It also enables novel applications, such as the curvature-based construction of perturbative black-hole deformations about Kerr. More fundamentally, our results suggest that generic perturbations of spacetimes perturbatively connected to a Petrov type D background may admit a local, order-by-order description in terms of the Weyl scalars $\Psi_{0,4}$ and the stress-energy tensor, even when the full, corrected spacetime is Petrov type I. We discuss this possibility later and will investigate its broader scope in future work.

\vspace{0.2cm}
\noindent\textit{Preliminaries---}Our starting point is the Einstein equations for the metric $g_{\mu\nu}$ in $3+1$ dimensions, which we write as
\begin{equation} \label{eq:Einstein_equation}
    [\mathcal{E}g]_{\mu\nu}= 8\pi\epsilon\, T_{\mu\nu}\,,
\end{equation}
where $\mathcal{E}$ is the Einstein operator, $T_{\mu\nu}$ is some stress-energy tensor, and $\epsilon$ is a small bookkeeping parameter (e.g.~$\epsilon$ could represent the smallness of a mass ratio or a dimensionless beyond-Einstein coupling), so the source is perturbative. For example, $T_{\mu\nu}$ could be the point-particle source of extreme mass-ratio inspirals, or the effective stress-energy tensor in effective-field-theory extensions of general relativity. To solve Eq.~\eqref{eq:Einstein_equation} perturbatively, we expand the metric as
\begin{align} \label{eq:metric_expansion}
    g_{\mu\nu}=g_{\mu\nu}^{(0)}+\sum_{n=1}\epsilon^n \, h_{\mu\nu}^{(n)}\,,
\end{align}
where $g_{\mu\nu}^{(0)}$ is the metric of the background spacetime and $h_{\mu\nu}^{(n)}$ is the $n$th order metric perturbation tensor. Since we focus on black hole perturbations, we take $g_{\mu\nu}^{(0)}$ to be the metric of a Ricci-flat Petrov type D black hole spacetime (e.g., a Kerr black hole). Linearizing Eq.~\eqref{eq:Einstein_equation} about $\epsilon$, one can derive two Teukolsky equations for the perturbed Weyl scalars $\Psi_{0,4}^{(1)}$: 
\begin{equation} \label{eq:Teukolsky_equation}
    H_{0,4}^{(0)}\Psi_{0,4}^{(1)}=\mathcal{S}_{0,4}^{(1)}\,,
\end{equation}
where $\Psi_{0}\equiv C_{lmlm}$ and $\Psi_{4}\equiv C_{n\bar{m}n\bar{m}}$ are projections of the Weyl tensor $C_{\mu\nu\alpha\beta}$ to the NP tetrad $e_a^{\mu}=\{l^{\mu},n^{\mu},m^{\mu},\bar{m}^{\mu}\}$. The tetrad vectors $m^{\mu}$ and its complex conjugate $\bar{m}^{\mu}$ span the 2-dimensional sector orthogonal to $l^{\mu}$ and $n^{\mu}$. The Teukolsky differential operators that act on $\Psi_{0,4}^{(1)}$ in general relativity are denoted $H_{0,4}^{(0)}$, while $\mathcal{S}_{0,4}^{(1)}$ are the sources driven by $T_{\mu\nu}^{(1)}$ and defined in \cite{Teukolsky:1973ha, Li:2022pcy}. For simplicity, we will drop the superscript of any terms at $\mathcal{O}(\epsilon^0)$ whenever a perturbative expansion is made. Our goal is to reconstruct the metric perturbation $h_{\mu\nu}^{(1)}$ from a perturbed $\Psi_{0}^{(1)}$ or $\Psi_{4}^{(1)}$ that solves Eq.~\eqref{eq:Teukolsky_equation}. 

In total, there are four real, $\{h_{ll}^{(1)}, h_{ln}^{(1)},h_{nn}^{(1)},h_{m\bar{m}}^{(1)}\}$, and three complex, $\{h_{lm}^{(1)},h_{nm}^{(1)},h_{mm}^{(1)}\}$, metric components in the NP basis; the other components are the complex conjugate of the former. Using the four degrees of coordinate freedom, we impose that
\begin{align} \label{eq:radiation_gauges_traceful}
    h_{\mu\nu}^{(1)}e_{a}^{\mu}=0\,,
\end{align}
where $e_{a}^{\mu}$ is either $l^{\mu}$ or $n^{\mu}$. More concretely, one can solve for the gauge vector $\xi^{\mu (1)}$, defined via $x^{\mu} \rightarrow x^{\mu}+\epsilon\,\xi^{\mu (1)}$, such that
\begin{equation}
    h_{\mu\nu}^{(1)}e_{a}^{\mu}\rightarrow
    \left[h_{\mu\nu}^{(1)}-2\xi_{(\mu;\nu)}^{(1)}\right]e_{a}^{\mu}=0\,.
\end{equation}
We work locally in smooth domains where this gauge can be imposed. For distributional sources, a globally smooth gauge vector needs not exist, and appropriate gauge patches or string prescriptions and source-matching conditions may be required \cite{Barack:2001ph}.

Unlike the traditional radiation gauges defined in Eq.~\eqref{eq:radiation_gauges_tracefree}, we do not impose the residual gauge condition $h^{(1)}=0$. As emphasized in \cite{Price:2006ke, Green:2019nam, Loutrel:2020wbw}, this additional condition is self-consistent only for restricted sources:
\begin{equation} \label{eq:stress_condition_traceful}
     T_{ll}^{(1)}=0\;\textrm{(IRG)}\quad\textrm{or}\quad
     T_{nn}^{(1)}=0\;\textrm{(ORG)}\,,
\end{equation}
where IRG and ORG stand for the ingoing [i.e., $e_{a}^{\mu}=l^{\mu}$ in Eq.~\eqref{eq:radiation_gauges_tracefree}] and outgoing [i.e., $e_{a}^{\mu}=n^{\mu}$ in Eq.~\eqref{eq:radiation_gauges_tracefree}] radiation gauges, respectively. We will instead show that the trace $h^{(1)}$ can be determined from a set of transport equations, and all the other metric components can be systematically solved from a set of commutation relations, Ricci identities, and Bianchi identities \cite{Chandrasekhar_1983, Loutrel:2020wbw, Ripley:2020xby}. To stay close to the procedures in \cite{Loutrel:2020wbw, Ripley:2020xby}, we will work with the \textit{traceful} ORG [i.e., $e_{n}^{\mu}=n^{\mu}$ in Eq.~\eqref{eq:radiation_gauges_traceful}], so the only nonzero metric components are 
\begin{equation} \label{eq:metric_ORG_traceful}
    \{h_{ll}^{(1)},h_{l\bar{m}}^{(1)},h_{\bar{m}\bar{m}}^{(1)},
    h_{m\bar{m}}^{(1)}\}
\end{equation}
and their complex conjugate; the procedure for the traceful IRG mirrors the former. Since $h_{ln}^{(1)}=0$ in this gauge, $h^{(1)}=2h_{m\bar{m}}^{(1)}$ in the mostly positive metric signature.

\begin{figure*}[htb]
\centering
\resizebox{1.6\columnwidth}{!}{%
\begin{tikzpicture}[
    scale=1,
    transform shape,
    >=stealth,
    font=\footnotesize,
    line width=0.5pt,
    eq/.style={font=\normalsize},
    lab/.style={font=\small, inner sep=2pt},
    data/.style={font=\small, inner sep=2pt}
]
\node[eq, align=center] (A) at (0,-1.6)
    {$H_4\Psi_4^{(1)}=\mathcal S_4^{(1)}$};
\node[eq, align=center] (L) at (0,0) {$\Phi_{22}^{(1)}$};

\node[eq, anchor=west] (B) at (4.1,-1.6) {$\Psi_4^{(1)}$};
\node[eq, anchor=west] (C) at (7.2,-1.6) {$\lambda^{(1)}$};
\node[eq, anchor=west] (D) at (10.3,-1.6)
    {$\textcolor{red}{h_{\bar m\bar m}^{(1)}}$};

\node[eq, anchor=west] (E) at (4.1,-3.6) {$\Psi_3^{(1)}$};
\node[eq, anchor=west] (F) at (7.2,-3.6) {$\pi^{(1)}$};
\node[eq, anchor=west] (G) at (10.3,-3.6)
    {$\textcolor{red}{h_{l\bar m}^{(1)}}
      \;\Rightarrow\;\alpha^{(1)},\beta^{(1)},\tau^{(1)}$};

\node[eq, anchor=west] (H) at (4.1,-5.6) {$\Psi_2^{(1)}$};
\node[eq, anchor=west] (I) at (7.2,-5.6)
    {$\textcolor{red}{h_{ll}^{(1)}}
      \;\Rightarrow\;\kappa^{(1)},\sigma^{(1)},
      \varepsilon^{(1)},\rho^{(1)},\Psi_0^{(1)},\Psi_1^{(1)}$};

\node[eq, anchor=west] (J) at (4.1,0) {$\mu^{(1)}$};
\node[eq, anchor=west] (K) at (7.2,0)
    {$\textcolor{red}{h_{m\bar m}^{(1)}}\;\Rightarrow\;\gamma^{(1)}$};

\draw[->] ($(A.east)+(0.35,0)$) -- ($(B.west)+(-0.25,0)$);

\draw[->] (L.east) --
    node[above,lab] {Eq.~\eqref{eq:mu_transport_final}}
    node[below,data] {$c_\mu$} (J.west);
\draw[->] (J.east) --
    node[above,lab] {Eq.~\eqref{eq:trace_transport}}
    node[below,data] {$c_{m\bar m}$} (K.west);

\draw[->] (B.east) --
    node[above,lab] {Eq.~\eqref{eq:lambda_transport}}
    node[below,data] {$c_\lambda$} (C.west);
\draw[->] (C.east) --
    node[above,lab] {Eq.~\eqref{eq:hmm_transport}}
    node[below,data] {$c_{\bar m\bar m}$} (D.west);

\draw[->] (B.south) --
    node[right,lab] {Eq.~\eqref{eq:psi3_transport}}
    node[left,data] {$c_{\Psi_3}$} (E.north);
\draw[->] (E.east) --
    node[above,lab] {Eq.~\eqref{eq:pi_transport}}
    node[below,data] {$c_\pi$} (F.west);
\draw[->] (F.east) --
    node[above,lab] {Eq.~\eqref{eq:hlm_transport}}
    node[below,data] {$c_{l\bar m}$} (G.west);

\draw[->] (E.south) --
    node[right,lab] {Eq.~\eqref{eq:psi2_transport}}
    node[left,data] {$c_{\Psi_2}$} (H.north);
\draw[->] (H.east) --
    node[above,lab] {Eq.~\eqref{eq:hll_transport}}
    node[below,data] {$c_{ll}$} (I.west);
\end{tikzpicture}
}
\caption{Schematic reconstruction procedure. The quantities at the endpoint of each single arrow $\rightarrow$ can be found by solving
the transport equation at this arrow. The quantities at the endpoint of each double arrow $\Rightarrow$ can be directly computed at the
corresponding reconstruction step. The homogeneous integration functions $c_f$ associated with each reconstructed field $f$ ($c_{ab}$ for $h_{ab}^{(1)}$) are defined on a cross-section transverse to $n^\mu$. Their admissible combinations satisfy the Einstein and Teukolsky--Bianchi compatibility conditions and are selected by the prescribed boundary and source-matching conditions, residual-gauge choices, and mass/angular-momentum completion data. In the traceful ORG, $h_{ln}^{(1)}=h_{nn}^{(1)}=h_{nm}^{(1)}=\nu^{(1)}=0$.}
\label{fig:reconstruction_flowchart}
\end{figure*}

\vspace{0.2cm}
\noindent\textit{Reconstruction of the trace---}To reconstruct $h_{\mu\nu}^{(1)}$, let us first use the remaining three (complex) degrees of tetrad-rotation freedom to fix the tetrad at $\mathcal{O}(\epsilon)$. One widely used choice in the ORG is:
\begin{equation} \label{eq:perturbed_tetrad}
    \begin{aligned}
        & l^{\mu(1)}=\frac{1}{2}h_{ll}^{(1)}n^{\mu}\,,\quad
        n^{\mu(1)}=0\,, \\
        & m^{\mu(1)}=h_{lm}^{(1)}n^{\mu}
        -\frac{1}{2}h_{m\bar{m}}^{(1)}m^{\mu}
        -\frac{1}{2}h_{mm}^{(1)}\bar{m}^{\mu}\,,
    \end{aligned}
\end{equation}
which only uses the orthogonality condition, three complex tetrad rotations, and the traceful ORG condition \cite{Chrzanowski:1975wv, Campanelli:1998jv, Loutrel:2020wbw}. For convenience, let us introduce the following shorthand notation for differential operators:
\begin{align} \label{eq:derivative-notation}
    D_{[a,b,c,d]}
    &=D+a \, \varepsilon+b \, \bar{\varepsilon}+c \, \rho+d \, \bar{\rho}\,, \nonumber\\
    \boldsymbol{\Delta}_{[a,b,c,d]}
    &=\boldsymbol{\Delta}+a \, \mu + b \, \bar{\mu}+ c \, \gamma+ d \,  \bar{\gamma}\,, \nonumber\\
    \delta_{[a,b,c,d]}
    &=\delta+a \, \bar{\alpha}+ \,  b\beta+c \,  \bar{\pi}+d \,  \tau\,, \nonumber\\
    \bar{\delta}_{[a,b,c,d]}
    &=\bar{\delta}+a \,  \alpha+b \,  \bar{\beta}+c \, \pi+d \,  \bar{\tau}\,,
\end{align}
where $\{D,\boldsymbol{\Delta},\delta,\bar{\delta}\}$ are directional derivatives along the null tetrad $\{l^{\mu},n^{\mu},m^{\mu},\bar{m}^{\mu}\}$, $\{a,b,c,d\}$ are integers, and the remaining quantities are spin coefficients. Our goal is to reconstruct the metric components in Eq.~\eqref{eq:metric_ORG_traceful} from the NP equations, which we provide in the supplemental material for completeness. For a review of the NP formalism, we refer the reader to \cite{Newman:1961qr, Chandrasekhar_1983, Pound:2019lzj, Loutrel:2020wbw, Li:2022pcy}.

To find $h_{m\bar{m}}^{(1)}$, let us focus on the perturbed spin coefficients $\mu^{(1)}$ and $\gamma^{(1)}$, as they only involve $h_{m\bar{m}}^{(1)}$ in the traceful ORG. Linearizing the commutation relations, one can express $\mu^{(1)}$ and $\gamma^{(1)}$ in terms of $h_{ab}^{(1)}$ [i.e., Eqs.~\eqref{eq:reconstruct_mu} and \eqref{eq:reconstruct_gamma}], which in the traceful ORG reduce to:
\begin{equation} \label{eq:reconstruct_mu_gamma_ORG} 
    \mu^{(1)}
    =\frac{1}{2}\boldsymbol{\Delta}_{[-1,1,0,0]}h_{m\bar{m}}^{(1)}\,,\quad
    \gamma^{(1)}
    =-\frac{1}{4}(\mu-\bar{\mu})h_{m\bar{m}}^{(1)}\,.   
\end{equation}
Since we work in the mostly positive metric signature, Eq.~\eqref{eq:perturbed_spin_coefs} differs from \cite{Loutrel:2020wbw} by an overall sign. To obtain $h_{m\bar{m}}^{(1)}$, we then need to solve for $\mu^{(1)}$ or $\gamma^{(1)}$. Linearizing the Ricci identity in Eq.~\eqref{eq:Ricci_Phi22}, we find
\begin{equation} \label{eq:mu_transport}
    \boldsymbol{\Delta}_{[2,0,1,1]}\mu^{(1)}
    +\mu^{(0)}\left(\gamma^{(1)}+\bar{\gamma}^{(1)}\right)
    +\Phi_{22}^{(1)}=0\,,
\end{equation}
where we have used that $\lambda^{(0)}=\nu^{(0)}=0$ for Petrov type D backgrounds, and $\boldsymbol{\Delta}^{(1)}=\nu^{(1)}=0$ [i.e., Eqs.~\eqref{eq:perturbed_tetrad} and \eqref{eq:reconstruct_nu}] in the traceful ORG. Furthermore, $\gamma^{(1)}$ is imaginary in the traceful ORG given Eq.~\eqref{eq:reconstruct_mu_gamma_ORG} and that $h_{m\bar{m}}^{(1)}$ is real, so the second term on the left-hand side of Eq.~\eqref{eq:mu_transport} vanishes. Thus, we find a transport equation for $\mu^{(1)}$ in terms of the NP Ricci scalar $\Phi_{22}^{(1)}$:
\begin{equation} \label{eq:mu_transport_final}
    \boldsymbol{\Delta}_{[2,0,1,1]}\mu^{(1)}
    =-\Phi_{22}^{(1)}\,.
\end{equation}
Solving for $\mu^{(1)}$ from Eq.~\eqref{eq:mu_transport_final}, one can then solve
\begin{equation} \label{eq:trace_transport}
    \boldsymbol{\Delta}_{[-1,1,0,0]}h_{m\bar{m}}^{(1)}
    =2\mu^{(1)}\,.
\end{equation}
Equation~\eqref{eq:trace_transport} then gives $h_{m\bar m}^{(1)}$, after which $\gamma^{(1)}$ follows from Eq.~\eqref{eq:reconstruct_gamma}. This trace equation is the step that removes the standard source restriction. A similar treatment of the trace was carried out in \cite{Suvorov:2019qow} for $f(R)$ gravity, and is thus extended here to generic sources. As shown in the supplemental material, combining Eqs.~\eqref{eq:mu_transport_final} and \eqref{eq:trace_transport} and applying the Geroch-Held-Penrose (GHP) prime operation \cite{Geroch:1973am} reproduces Eq.~(56) of the corrector-tensor approach in \cite{Green:2019nam}. This overlap reveals a deeper connection between these two approaches, but the equation serves different roles. In \cite{Green:2019nam}, it determines the corrector tensor's trace so that the residual source is compatible with a Hertz-potential reconstruction in a tracefree radiation gauge, whereas here it forms the trace branch of our reconstruction hierarchy in a traceful radiation gauge.

\vspace{0.2cm}
\noindent\textit{Reconstruction of other components---}The reconstruction of the remaining components in Eq.~\eqref{eq:metric_ORG_traceful} follows the hierarchical structure discussed in \cite{Loutrel:2020wbw}. Let us first linearize the Ricci identity in Eq.~\eqref{eq:Ricci_Psi4} to obtain a transport equation for $\lambda^{(1)}$ in terms of $\Psi_4^{(1)}$, i.e.,
\begin{equation} \label{eq:lambda_transport}
    \boldsymbol{\Delta}_{[1,1,3,-1]}\lambda^{(1)}=-\Psi_4^{(1)}\,,
\end{equation}
where we again use that $\nu^{(1)}=0$ in the traceful ORG. Using Eq.~\eqref{eq:reconstruct_lambda}, which relates $\lambda^{(1)}$ to $h_{ab}^{(1)}$, we further find
\begin{equation} \label{eq:hmm_transport}
    \boldsymbol{\Delta}_{[-1,1,2,-2]}h_{\bar{m}\bar{m}}^{(1)}
    =2\lambda^{(1)}\,.
\end{equation}
Solving $\lambda^{(1)}$ first from Eq.~\eqref{eq:lambda_transport}, one can then obtain $h_{\bar{m}\bar{m}}^{(1)}$ from Eq.~\eqref{eq:hmm_transport}. Our Eqs.~\eqref{eq:lambda_transport} and \eqref{eq:hmm_transport} align with those in \cite{Loutrel:2020wbw}, but generalized to traceful ORG.

Next, to find $h_{l\bar{m}}^{(1)}$, we follow \cite{Loutrel:2020wbw} to solve for $\Psi_3^{(1)}$ first by linearizing the Bianchi identity in Eq.~\eqref{eq:Bianchi_Delta_Psi3}:
\begin{equation} \label{eq:psi3_transport}
    \boldsymbol{\Delta}_{[4,0,2,0]}\Psi_{3}^{(1)}
    =\delta_{[0,4,0,-1]}\Psi_{4}^{(1)}+\mathcal{R}_h^{(1)}\,,
\end{equation}
where $\mathcal{R}_h$ is defined in Eq.~\eqref{eq:Bianchi_Delta_Psi3_source} in terms of the NP Ricci scalars $\Phi_{ij}$. Linearizing the Ricci identity in Eq.~\eqref{eq:Ricci_Psi3_Phi21}, we find a transport equation for $\pi^{(1)}$:
\begin{align} \label{eq:pi_transport}
    \boldsymbol{\Delta}_{[0,0,1,-1]}\pi^{(1)}
    =& \;-\Psi_3^{(1)}-\Phi_{21}^{(1)}-\frac{1}{2}(\pi+\bar{\tau})
    \boldsymbol{\Delta}_{[0,1,0,0]}h_{m\bar{m}}^{(1)} \nonumber\\
    & \;-\frac{1}{2}(\bar{\pi}+\tau)
    \boldsymbol{\Delta}_{[0,1,2,-2]}h_{\bar{m}\bar{m}}^{(1)} \nonumber\\
    & \;+\frac{1}{2}(\mu-\bar{\mu})\pi h_{m\bar{m}}^{(1)}\,,
\end{align}
where we have used Eqs.~\eqref{eq:mu_transport_final}, \eqref{eq:hmm_transport}, and that
\begin{equation}
    \pi^{(1)}+\bar{\tau}^{(1)}
    =\frac{1}{2}(\pi+\bar{\tau})h_{m\bar{m}}^{(1)}
    +\frac{1}{2}(\bar{\pi}+\tau)h_{\bar{m}\bar{m}}^{(1)}\,. 
\end{equation}
From the relation between $\pi^{(1)}$ and $h_{ab}^{(1)}$ in Eq.~\eqref{eq:reconstruct_pi}, we finally find a transport equation for $h_{l\bar{m}}^{(1)}$,
\begin{equation} \label{eq:hlm_transport}
    \boldsymbol{\Delta}_{[0,1,0,-2]}h_{l\bar{m}}^{(1)}
    =2\pi^{(1)}-\bar{\tau}h_{m\bar{m}}^{(1)}
    -\tau h_{\bar{m}\bar{m}}^{(1)}\,.
\end{equation}
Using Eqs.~\eqref{eq:reconstruct_alpha}, \eqref{eq:reconstruct_beta}, and \eqref{eq:reconstruct_tau}, one can now compute $\alpha^{(1)}$, $\beta^{(1)}$, and $\tau^{(1)}$ with the known metric components.

Finally, to find $h_{ll}^{(1)}$, we first reconstruct $\Psi_2^{(1)}$ using the Bianchi identity in Eq.~\eqref{eq:Bianchi_Delta_Psi2}, which linearizes to
\begin{equation} \label{eq:psi2_transport}
    \boldsymbol{\Delta}_{[3,0,0,0]}\Psi_2^{(1)}
    =-3\Psi_2\mu^{(1)}
    +\delta_{[0,2,0,-2]}\Psi_3^{(1)}+\mathcal{R}_g^{(1)}\,.
\end{equation}
We then linearize the Ricci identity in Eq.~\eqref{eq:Ricci_Psi2_Lambda},
\begin{equation} \label{eq:Ricci_Psi2_Lambda_linearize}
\begin{aligned}
    D_{[1,1,0,-1]}^{(1)}\mu
    =& \;-D_{[1,1,0,-1]}\mu^{(1)}
    +\delta_{[-1,1,1,0]}^{(1)}\pi \\
    & \;+\delta_{[-1,1,1,0]}\pi^{(1)}
    +\Psi_2^{(1)}+2\Lambda^{(1)}\,.  
\end{aligned}
\end{equation}
After using the expressions for $\varepsilon^{(1)}$ and $\rho^{(1)}$ in terms of $h_{ab}^{(1)}$ in Eqs.~\eqref{eq:reconstruct_epsilon} and \eqref{eq:reconstruct_rho}, we reduce $D_{[1,1,0,-1]}^{(1)}\mu$ to 
\begin{equation}
\begin{aligned}
   D_{[1,1,0,-1]}^{(1)}\mu
   =& \;\frac{1}{2}\mu\Big(\boldsymbol{\Delta}_{[-1,1,-2,-2]}h_{ll}^{(1)}
   -\bar{\delta}_{[-2,0,-1,0]}h_{lm}^{(1)} \\
   & \;+\delta_{[-2,0,1,2]}h_{l\bar{m}}^{(1)}
   +D_{[0,0,-1,1]}h_{m\bar{m}}^{(1)}\Big)\,, 
\end{aligned} 
\end{equation}
where we have used that $\boldsymbol{\Delta}\mu=-\mu^2-\mu(\gamma+\bar{\gamma})$ at $\mathcal{O}(\epsilon^0)$. Thus, $h_{ll}^{(1)}$ satisfies the transport equation
\begin{equation} \label{eq:hll_transport}
\begin{aligned}
    \boldsymbol{\Delta}_{[-1,1,-2,-2]}h_{ll}^{(1)}
    =& \;\bar{\delta}_{[-2,0,-1,0]}h_{lm}^{(1)}
    -\delta_{[-2,0,1,2]}h_{l\bar{m}}^{(1)} \\
    & \;-D_{[0,0,-1,1]}h_{m\bar{m}}^{(1)} 
    +\frac{2}{\mu}\mathcal{S}_{\mu}^{(1)}\,,
\end{aligned}
\end{equation}
where $\mathcal{S}_{\mu}^{(1)}$ is the right-hand side of Eq.~\eqref{eq:Ricci_Psi2_Lambda_linearize}. Equation~\eqref{eq:hll_transport} assumes $\mu\neq0$. If $\mu$ vanishes only at a boundary, the equation can be rewritten for $\tilde h_{ll}^{(1)}=\mu h_{ll}^{(1)}$ \cite{Ripley:2020xby}, removing the explicit division by $\mu$; regularity must still be assessed in a horizon-regular chart and tetrad. If $\mu$ vanishes identically in special cases \cite{DeGroote:2009ah}, this reconstruction route becomes degenerate, but an alternative route that follows Eq.~(33) of \cite{Loutrel:2020wbw} can be used. Both prescriptions and their compatibility requirements are given in the Supplemental Material.

Equations~\eqref{eq:mu_transport_final}--\eqref{eq:pi_transport}, \eqref{eq:hlm_transport}, \eqref{eq:psi2_transport}, and \eqref{eq:hll_transport} therefore provide a closed hierarchical reconstruction of $h^{(1)}_{ab}$ for a generic perturbative source in the traceful ORG, as summarized in Fig.~\ref{fig:reconstruction_flowchart}. At each step, the source is either a NP Ricci scalar determined directly by $T_{\mu\nu}^{(1)}$, which involves no geometric quantities at $\mathcal{O}(\epsilon)$, or a quantity reconstructed in an earlier step. After harmonic decomposition, each equation reduces to a first-order radial transport equation along the ingoing principal null direction. This reconstruction procedure also has an intrinsic spin-weight structure: $\Psi_4^{(1)}$ reconstructs the spin-$2$ $\lambda^{(1)}$ and $h_{\bar{m}\bar{m}}^{(1)}$ first, with the spin-$1$ and spin-$0$ sectors following. Compared to \cite{Loutrel:2020wbw}, our main innovative step is an additional spin-$0$ branch, where $h_{m\bar{m}}^{(1)}$ is reconstructed directly from $T_{nn}^{(1)}$ through Eqs.~\eqref{eq:mu_transport_final} and \eqref{eq:trace_transport}; the remaining equations follow \cite{Loutrel:2020wbw} with extra source terms involving $h_{m\bar m}^{(1)}$. Our result also generalizes the one in \cite{Suvorov:2019qow}, which was specialized to $f(R)$ gravity and formulated in the traceful IRG. 

Although the hierarchy uses only a subset of the NP equations, we prove in the supplemental material that for a conserved source, an input $\Psi_4^{(1)}$ satisfying Eq.~\eqref{eq:Teukolsky_equation}, and compatible integration data, solving this subset yields a metric satisfying the \textit{complete set} of linearized Einstein equations throughout the smooth reconstruction domain. For a retarded solution, the integration constants are fixed by regularity at the future horizon, no incoming radiation and source-compatible asymptotic behavior at infinity, matching or jump conditions across discontinuities, physical completion data such as mass and angular-momentum shifts, and a residual-gauge prescription. These homogeneous functions associated with each individual transport equation are shown in Fig.~\ref{fig:reconstruction_flowchart}.

As a simple demonstration of our approach, we apply it to a Schwarzschild black hole surrounded by a thin, static, spherical shell in the supplemental material. In this example, the spacetime is Schwarzschild on both sides of the thin shell, so the radiative Weyl scalars $\Psi_{0,4}^{(1)}$ vanish, and junction conditions \cite{Israel:1966rt, Poisson:2009pwt} ensure that the metric is continuous across the shell when supported by a stress-energy tensor in the perfect-fluid form \cite{Frauendiener:1990nao, Laeuger:2025zgb}. When the shell's mass is tiny, it induces a monopolar perturbation that shifts the exterior ADM mass, so the spin-2 and spin-1 sectors of $h_{\mu\nu}^{(1)}$ vanish by spherical symmetry. The spin-0 components $h_{m\bar m}^{(1)}$ and $h_{ll}^{(1)}$ are then reconstructed from the shell's stress-energy tensor together with the boundary, completion, and matching data via Eqs.~\eqref{eq:mu_transport_final}, \eqref{eq:trace_transport}, \eqref{eq:psi2_transport}, and \eqref{eq:hll_transport}. Similar to the known result in \cite{Frauendiener:1990nao}, our reconstructed metric is free of distributional singularities and continuous across the shell. In the supplemental material, we show explicitly that the perturbed metric we reconstruct matches the linearization of the exact solution found in \cite{Frauendiener:1990nao}, when transformed to the same gauge.

\vspace{0.2cm}
\noindent\textit{Comparison to other approaches and future directions---}
Having developed a metric reconstruction procedure for generic sources of perturbations to Petrov type D background spacetimes, we now discuss its relation to existing approaches, potential challenges, and future applications.

Once the gauge, boundary, and completion data are fixed, different valid reconstruction methods should recover the same physical metric. Our framework is nevertheless procedurally distinct from existing CCK-like approaches \cite{Lousto:2002em, Ori:2002uv, Lousto:2005xu, Green:2019nam, Toomani:2021jlo, Bourg:2024vre, Hollands:2024iqp, Dolan:2021ijg, Dolan:2023enf, Nasipak:2025tby}. When the physical Weyl scalar is taken as input, these CCK-like approaches require an additional curvature-to-Hertz inversion. The inversion is algebraic mode by mode for non-static vacuum perturbations through the Teukolsky-Starobinsky identity \cite{Teukolsky:1974yv, Starobinsky:1973aij, Lousto:2002em, Ori:2002uv, vandeMeent:2017bcc}, whereas source regions require solving the fourth-order relation with suitable matching data \cite{Green:2019nam, Toomani:2021jlo, Bourg:2024vre, Hollands:2024iqp} or separately treating the static modes after generalizing the Teukolsky-Starobinsky identity \cite{Aksteiner:2016pjt, Dolan:2021ijg, Dolan:2023enf, Wardell:2024yoi}. Source-supported completion sectors can also be recovered self-consistently, but are typically organized separately from the radiative Hertz sector.

By contrast, once the physical $\Psi_{0,4}^{(1)}$ is known, our method reconstructs the total metric through first-order NP transport equations, without a Hertz potential, curvature-to-Hertz inversion, time antiderivative, or division by the Fourier frequency. Radiative, static, and completion sectors are treated within the same hierarchy. In particular, monopolar (e.g., mass shift) and dipolar (e.g., angular-momentum shift) completion data, although invisible to $\Psi_{0,4}^{(1)}$, enter through the homogeneous and boundary data of the lower-spin transport equations, such as the ones for the spin-1 ($h_{l\bar{m}}^{(1)}$) and spin-0 ($h_{m\bar m}^{(1)}$ and $h_{ll}^{(1)}$) components. The thin-shell example illustrates this explicitly: despite the fact that $\Psi_{0,4}^{(1)}=0$, the hierarchy reconstructs the nonzero spin-0 mass-shift sector. This framework may be advantageous for time-domain, zero- or low-frequency, and iterative effective-source calculations, although demonstrating these advantages quantitatively remains an important direction for future work.

An important issue for EMRI applications is that radiation-gauge perturbations generally contain string-like singularities extending from the particle or discontinuities across a surface through it \cite{Barack:2001ph, Ori:2002uv, Pound:2013faa, Green:2019nam, Dolan:2023enf}. Relaxing the tracefree condition does not remove this problem, as Barack and Ori showed that even a static particle in flat spacetime can require a singular radiation-gauge vector \cite{Barack:2001ph}. By contrast, in Lorenz gauge, the point-particle singularity is confined to the worldline and has a standard local structure \cite{Pound:2013faa, Dolan:2023enf}. On a Ricci-flat background, the transformation to the Lorenz-gauge metric $h^{\mathrm L}_{ab}$ is determined by $h^{\mathrm L}_{ab}
=h^{\mathrm{ORG}}_{ab}-2\nabla_{(a}\xi_{b)}$, where $\Box\xi_a=\nabla^b\bar h^{\mathrm{ORG}}_{ab}$ and $\bar h^{\mathrm{ORG}}_{ab}$ is the trace reversal of $h^{\mathrm{ORG}}_{ab}$. Because the source $\nabla^b\bar h^{\mathrm{ORG}}_{ab}$ is a fixed linear functional of the reconstructed radiation-gauge metric, its tetrad projection, the mode-decomposed gauge-vector equations, and their boundary and matching conditions can be formulated in advance. The numerical radiation-gauge metric then supplies the inhomogeneous source terms. Existing transformations from tracefree radiation gauges to Lorenz gauge provide a useful template \cite{Dolan:2021ijg, Dolan:2023enf}, although the traceful sourced case requires an adapted gauge-vector solve and, for a point particle, suitable puncture or patching conditions. 

The framework developed here opens many directions for future work. For the modeling of extreme mass-ratio inspirals, our framework can provide the first-order metric perturbation needed for second-order self-force calculations, which are essential for waveform accuracy requirements of space-based detectors \cite{Aksteiner:2016pjt, Wardell:2024yoi, Green:2019nam, Toomani:2021jlo, Bourg:2024vre, Hollands:2024iqp, Dolan:2021ijg, Dolan:2023enf, Nasipak:2025tby, Loutrel:2020wbw, Ripley:2020xby}. Because our procedure applies to generic perturbations and treats static and completion modes within the same hierarchy, it also complements existing CCK-like methods and motivates hybrid reconstruction schemes. In future work, we will apply this approach to point-particle sources around Schwarzschild and Kerr black holes to benchmark our results against known results for circular, equatorial orbits \cite{Barack:2005nr, Barack:2007tm, Bourg:2024vre, Dolan:2023enf}.

Another novel application is a curvature-based route to stationary deformations of black hole geometries in non-vacuum environments or beyond-Einstein theories. In this setting, $T_{\mu\nu}$ in Eq.~\eqref{eq:Einstein_equation} represents a stationary effective source, such as a matter stress-energy tensor of an astrophysical environment, or higher-curvature corrections of a beyond-Einstein theory, while $\epsilon$ controls the strength of the deformation. For deformations whose corrected background spacetime remains Petrov type D, the problem closely resembles reconstructing completion modes, as illustrated by the thin-shell example. For deformations whose corrected background spacetime is Petrov type I, such as rotating black holes in dynamical Chern-Simons gravity \cite{Owen:2021eez}, the stationary $\Psi_{0,4}^{(1)}$ are nonzero and gauge-invariant quantities (similar to dynamical perturbations) that carry higher-order multipole information, which is then converted into metric components through the transport equations. Thus, the calculation reduces to solving a sourced stationary Teukolsky equation followed by first-order transport equations, which provides a useful complement to existing approaches for solving for the spacetime of non-vacuum and beyond-Einstein black holes \cite{Fernandes:2022gde, Garcia:2023ntf, Lam:2025elw, Lam:2025fzi, Fernandes:2025osu}.

More broadly, the reconstruction procedure developed here extends naturally to nonlinear orders and multi-parameter expansions. As in~\cite{Green:2019nam}, the principal parts of the transport equations remain unchanged order by order; higher-order effects therefore enter only through effective sources built from lower-order perturbations. This extension provides a metric-level step that complements the nonlinear modified Teukolsky formalism of \cite{Li:2022pcy}: the modified Teukolsky formalism gives decoupled, separable equations for $\Psi_{0,4}$ at each order in the expansion parameters, while the hierarchy derived here reconstructs the corresponding metric perturbation and hence the sources at subsequent orders. Together, these results provide a systematic curvature-based framework for studying nonlinear black hole perturbations in vacuum general relativity, non-vacuum environments, and beyond-Einstein theories. At a deeper level, our work suggests that, once the relevant global charges and boundary data are specified, perturbations of spacetimes perturbatively connected to Petrov type D backgrounds may admit a local description in terms of a single radiative Weyl scalar and the stress-energy tensor. Establishing the scope of this statement is an important direction for future work. Whether perturbations about a finite, generic Petrov type I background admit a perturbative description with the framework introduced here remains an open problem. 

\acknowledgments
\vspace{0.2cm}
\noindent\textit{Acknowledgments---}
    We are grateful to Colin Weller and Yanbei Chen for insightful discussions and constructive comments on the note that led to this work. We also thank Leor Barack, Roman Berens, Yanbei Chen, Takahiro Tanaka, Pratik Wagle, and Colin Weller for comments and suggestions on the draft, and Samuel Gralla, Adam Pound, and Barry Wardell for useful discussions. D.~L. and N.~Y. acknowledge support from the Simons Foundation (via Award No. 896696), the Simons Foundation International (via Grant No. SFI-MPS-BH-00012593-01), and the NSF (via Grants No.~PHY-2512423).

\clearpage
\appendix

\onecolumngrid
\section{Supplemental Material: Summary of NP equations}

We here provide the fundamental equations in the NP formalism, which include four commutation relations, eighteen Ricci identities, and eleven Bianchi identities. For a review of the NP formalism and the definition of all the NP quantities, we refer the reader to \cite{Newman:1961qr, Chandrasekhar_1983, Pound:2019lzj, Loutrel:2020wbw, Li:2022pcy}. 

In this work, we choose the mostly positive signature $(-,+,+,+)$, so the tetrad $\{l^{\mu},n^{\mu},m^{\mu},\bar{m}^{\mu}\}$ satisfies the orthogonality relations $l^\mu n_\mu=-1$, $m^\mu\bar{m}_\mu=1$, while the inner products of other tetrad vectors are zero. These tetrad vectors define four directional derivatives $\{D,\boldsymbol{\Delta},\delta,\bar{\delta}\}$, respectively, the latter of which satisfy the following commutation relations: 
\begin{subequations} \label{eq:commutation_relation}
\begin{align}
    [\boldsymbol{\Delta},D]
    &=(\gamma+\bar{\gamma})D
    +(\varepsilon+\bar{\varepsilon})\boldsymbol{\Delta}
    -(\bar{\tau}+\pi)\delta
    -(\tau+\bar{\pi})\bar{\delta}\,, \\
    [\delta,D]
    &=(\bar{\alpha}+\beta-\bar{\pi})D
    +\kappa\boldsymbol{\Delta}
    -(\bar{\rho}+\varepsilon-\bar{\varepsilon})\delta
    -\sigma\bar{\delta}\,, \\
    [\delta,\boldsymbol{\Delta}]
    &=-\bar{\nu}D
    +(\tau-\bar{\alpha}-\beta)\boldsymbol{\Delta}
    +(\mu-\gamma+\bar{\gamma})\delta
    +\bar{\lambda}\bar{\delta}\,, \\
    [\bar{\delta},\delta]
    &=(\bar{\mu}-\mu)D
    +(\bar{\rho}-\rho)\boldsymbol{\Delta}
    +(\alpha-\bar{\beta})\delta
    +(\beta-\bar{\alpha})\bar{\delta}\,,
\end{align}
\end{subequations}
which essentially define the twelve spin coefficients $\{\kappa,\sigma,\lambda,\nu,\varepsilon, \rho,\mu,\gamma,\alpha,\beta,\pi,\tau\}$ \cite{Chandrasekhar_1983}. 

In the NP formalism, the Riemann tensor is decomposed into five Weyl scalars $\{\Psi_0,\Psi_1,\Psi_2,\Psi_3,\Psi_4\}$ and ten NP Ricci scalars, including $\Phi_{ij}$ with $i,j\in\{0,1,2\}$ and $\Lambda$. These components of the Riemann tensor are related to the spin coefficients via eighteen Ricci identities:
\begin{subequations} \label{eq:Ricci_identities}
\begin{align}
    & D_{[-1,-1,-1,0]}\rho-\bar{\delta}_{[-3,-1,1,0]}\kappa
    -\sigma\bar{\sigma}+\bar{\kappa}\tau=\Phi_{00}\,, \\
    & D_{[-3,1,-1,-1]}\sigma-\delta_{[-1,-3,1,-1]}\kappa
    =\Psi_0\,, \\
    & D_{[-1,1,-1,0]}\tau-\boldsymbol{\Delta}_{[0,0,-3,-1]}\kappa
    -\rho\bar{\pi}-\sigma(\bar{\tau}+\pi)
    =\Psi_1+\Phi_{01}\,, \\
    & D_{[2,-1,-1,0]}\alpha-\bar{\delta}_{[0,-1,1,0]}\varepsilon
    -\beta\bar{\sigma}+\kappa\lambda
    +\bar{\kappa}\gamma-\pi\rho=\Phi_{10}\,, \\
    & D_{[0,1,0,-1]}\beta-\delta_{[-1,0,1,0]}\varepsilon
    -\sigma(\alpha+\pi)+\kappa(\mu+\gamma)=\Psi_1\,, \\
    & D_{[1,1,0,0]}\gamma
    -\boldsymbol{\Delta}_{[0,0,-1,-1]}\varepsilon
    -\alpha(\tau+\bar{\pi})-\beta(\bar{\tau}+\pi)
    -\tau\pi+\nu\kappa=\Psi_2+\Phi_{11}-\Lambda\,, \label{eq:Ricci_Phi11_Lambda} \\
    & D_{[3,-1,-1,0]}\lambda-\bar{\delta}_{[1,-1,1,0]}\pi
    -\bar{\sigma}\mu+\nu\bar{\kappa}=\Phi_{20}\,, \\
    & D_{[1,1,0,-1]}\mu-\delta_{[-1,1,1,0]}\pi
    -\sigma\lambda+\nu\kappa=\Psi_2+2\Lambda\,,  \label{eq:Ricci_Psi2_Lambda} \\
    & D_{[3,1,0,0]}\nu-\boldsymbol{\Delta}_{[1,0,1,-1]}\pi
    -\mu\bar{\tau}-\lambda(\bar{\pi}+\tau)
    =\Psi_3+\Phi_{21}\,, \label{eq:Ricci_Psi3_Phi21} \\
    & \boldsymbol{\Delta}_{[1,1,3,-1]}\lambda
    -\bar{\delta}_{[3,1,1,-1]}\nu=-\Psi_4\,, \label{eq:Ricci_Psi4} \\
    & \delta_{[-1,-1,0,-1]}\rho-\bar{\delta}_{[-3,1,0,0]}\sigma
    +\tau\bar{\rho}-\kappa(\mu-\bar{\mu})=-\Psi_1+\Phi_{01}\,, \\
    & \delta_{[-1,1,0,0]}\alpha-\bar{\delta}_{[-1,1,0,0]}\beta
    -\mu\rho+\lambda\sigma-\gamma(\rho-\bar{\rho})
    -\varepsilon(\mu-\bar{\mu})=-\Psi_2+\Phi_{11}+\Lambda\,, \\
    & \delta_{[-1,3,0,0]}\lambda-\bar{\delta}_{[1,1,1,0]}\mu
    -\nu(\rho-\bar{\rho})+\pi\bar{\mu}=-\Psi_3+\Phi_{21}\,, \\
    & \delta_{[1,3,0,-1]}\nu-\boldsymbol{\Delta}_{[1,0,1,1]}\mu
    -\lambda\bar{\lambda}+\bar{\nu}\pi=\Phi_{22}\,, \label{eq:Ricci_Phi22} \\
    & \delta_{[1,1,0,-1]}\gamma-\boldsymbol{\Delta}_{[1,0,-1,1]}\beta
    -\mu\tau+\sigma\nu+\varepsilon\bar{\nu}
    -\alpha\bar{\lambda}=\Phi_{12}\,, \\
    & \delta_{[1,-1,0,-1]}\tau
    -\boldsymbol{\Delta}_{[1,0,-3,1]}\sigma
    -\lambda\bar{\rho}+\kappa\bar{\nu}=\Phi_{02}\,, \\
    & \boldsymbol{\Delta}_{[0,1,-1,-1]}\rho
    -\bar{\delta}_{[-1,1,0,-1]}\tau
    -\sigma\lambda-\nu\kappa=-\Psi_2-2\Lambda\,, \\
    & \boldsymbol{\Delta}_{[0,1,0,-1]}\alpha
    -\bar{\delta}_{[0,1,0,-1]}\gamma
    -\nu(\rho+\varepsilon)+\lambda(\tau+\beta)=-\Psi_3\,,
\end{align}
\end{subequations}
which are the tetrad projections of the following identity to the NP basis \cite{Chandrasekhar_1983}:
\begin{equation} \label{eq:Ricci_identity}
    R_{abcd}
    =-\gamma_{abc,d}+\gamma_{abd,c}
    +\gamma_{baf}\left[\gamma_{c}{ }^{f}{ }_{d}
    -\gamma_{d}{ }^{f}{ }_{c}\right] 
    +\gamma_{fac}\gamma_{b}{ }^{f}{ }_{d}
    -\gamma_{fad}\gamma_{b}{ }^{f}{ }_{c}\,,
\end{equation}
after decomposing the Riemann tensor into the Weyl tensor, Ricci tensor, and Ricci scalar, where $\gamma_{abc}$ are the spin coefficients. The contraction of Eq.~\eqref{eq:Ricci_identity} onto the tetrad essentially gives the Einstein equations, once the equations are put on shell, i.e., once we set $R_{\mu\nu}=8\pi\left(T_{\mu\nu}-\frac{1}{2}g_{\mu\nu}T\right)$.

Finally, after projecting the Bianchi identity 
\begin{equation}
    R_{ab[cd;e]}=0
\end{equation}
to the NP tetrad, one finds eleven Bianchi identities, with eight being the transport equations of the Weyl scalars:
\begin{subequations} \label{eq:Bianchi_identities}
\begin{align}
    & \bar{\delta}_{[-4,0,1,0]}\Psi_0-D_{[-2,0,-4,0]}\Psi_1
    -3\kappa\Psi_2=\mathcal{R}_a\,, \\
    & \bar{\delta}_{[-2,0,2,0]}\Psi_1-D_{[0,0,-3,0]}\Psi_2
    -\lambda\Psi_0-2\kappa\Psi_3=-\mathcal{R}_b\,, \\
    & \bar{\delta}_{[0,0,3,0]}\Psi_2-D_{[2,0,-2,0]}\Psi_3
    -2\lambda\Psi_1-\kappa\Psi_4=\mathcal{R}_c\,, \\
    & \bar{\delta}_{[2,0,4,0]}\Psi_3-D_{[4,0,-1,0]}\Psi_4
    -3\lambda\Psi_2=-\mathcal{R}_d\,, \label{eq:Bianchi_bardelta_Psi3}\\
    & \boldsymbol{\Delta}_{[1,0,-4,0]}\Psi_0
    -\delta_{[0,-2,0,-4]}\Psi_1
    -3\sigma\Psi_2=\mathcal{R}_e\,, \\
    & \boldsymbol{\Delta}_{[2,0,-2,0]}\Psi_1
    -\delta_{[0,0,0,-3]}\Psi_2
    -\nu\Psi_0-2\sigma\Psi_3=\mathcal{R}_f\,, \\
    & \boldsymbol{\Delta}_{[3,0,0,0]}\Psi_2
    -\delta_{[0,2,0,-2]}\Psi_3
    -2\nu\Psi_1-\sigma\Psi_4=\mathcal{R}_g\,, \label{eq:Bianchi_Delta_Psi2} \\
    & \boldsymbol{\Delta}_{[4,0,2,0]}\Psi_3
    -\delta_{[0,4,0,-1]}\Psi_4
    -3\nu\Psi_2=\mathcal{R}_h\,, \label{eq:Bianchi_Delta_Psi3}
\end{align}
\end{subequations}
where the source terms $\mathcal{R}_{a},\cdots,\mathcal{R}_{h}$ are defined as
\begin{subequations} \label{eq:Bianchi_identities_source}
\begin{align}
    & \mathcal{R}_a
    =\delta_{[-2,-2,1,0]}\Phi_{00}-D_{[-2,0,0,-2]}\Phi_{01}
    +2\sigma\Phi_{10}-2\kappa\Phi_{11}-\bar{\kappa}\Phi_{02}\,, \\
    & \mathcal{R}_b
    =\bar{\delta}_{[-2,0,0,-2]}\Phi_{01}
    -\boldsymbol{\Delta}_{[1,0,-2,-2]}\Phi_{00}
    -2D\Lambda-2\tau\Phi_{10}
    +2\rho\Phi_{11}+\bar{\sigma}\Phi_{02}\,, \\
    & \mathcal{R}_c
    =\delta_{[-2,2,1,0]}\Phi_{20}-D_{[2,0,0,-2]}\Phi_{21}
    -2\bar{\delta}\Lambda-2\mu\Phi_{10}
    +2\pi\Phi_{11}-\bar{\kappa}\Phi_{22}\,, \\
    & \mathcal{R}_d
    =\bar{\delta}_{[2,0,0,-2]}\Phi_{21}
    -\boldsymbol{\Delta}_{[0,1,2,-2]}\Phi_{20}
    +2\nu\Phi_{10}+\bar{\sigma}\Phi_{22}-2\lambda\Phi_{11}\,, \\
    & \mathcal{R}_e
    =\delta_{[0,-2,2,0]}\Phi_{01}-D_{[-2,2,0,-1]}\Phi_{02}
    -2\kappa\Phi_{12}-\bar{\lambda}\Phi_{00}+2\sigma\Phi_{11}\,, \\
    & \mathcal{R}_f
    =\boldsymbol{\Delta}_{[0,2,-2,0]}\Phi_{01}
    -\bar{\delta}_{[-2,2,0,-1]}\Phi_{02}
    +2\delta\Lambda+2\tau\Phi_{11}
    -2\rho\Phi_{12}-\bar{\nu}\Phi_{00}\,, \\
    & \mathcal{R}_g
    =\delta_{[0,2,2,0]}\Phi_{21}-D_{[2,2,0,-1]}\Phi_{22}
    -2\boldsymbol{\Delta}\Lambda-2\mu\Phi_{11}
    +2\pi\Phi_{12}-\bar{\lambda}\Phi_{20}
    \,,  
    \label{eq:Bianchi_Delta_Psi2_source} \\
    & \mathcal{R}_h
    =\boldsymbol{\Delta}_{[0,2,2,0]}\Phi_{21}
    -\bar{\delta}_{[2,2,0,-1]}\Phi_{22}
    -2\nu\Phi_{11}-\bar{\nu}\Phi_{20}+2\lambda\Phi_{12}\,.
    \label{eq:Bianchi_Delta_Psi3_source}
\end{align}
\end{subequations}
The remaining three impose the divergence-free condition of the Einstein tensor and only involve the NP Ricci scalars: 
\begin{subequations} \label{eq:Bianchi_div} 
\begin{align}
    & \bar{\delta}_{[-2,0,1,-2]}\Phi_{01}
    +\delta_{[-2,0,1,-2]}\Phi_{10}
    -D_{[0,0,-2,-2]}\Phi_{11}
    -\boldsymbol{\Delta}_{[1,1,-2,-2]}\Phi_{00}
    -\bar{\kappa}\Phi_{12}-\kappa\Phi_{21}
    +\bar{\sigma}\Phi_{02}+\sigma\Phi_{20}
    -3D\Lambda=0\,, \label{eq:Bianchi_div1} \\
    & \bar{\delta}_{[0,2,2,-1]}\Phi_{12}
    +\delta_{[0,2,2,-1]}\Phi_{21}
    -\boldsymbol{\Delta}_{[2,2,0,0]}\Phi_{11}
    -D_{[2,2,-1,-1]}\Phi_{22}
    +\nu\Phi_{01}+\bar{\nu}\Phi_{10}
    -\lambda\Phi_{02}-\bar{\lambda}\Phi_{20}
    -3\boldsymbol{\Delta}\Lambda=0\,, \label{eq:Bianchi_div2} \\
    & \delta_{[0,0,2,-2]}\Phi_{11}
    -D_{[0,2,-2,-1]}\Phi_{12}
    -\boldsymbol{\Delta}_{[1,2,-2,0]}\Phi_{01}
    +\bar{\delta}_{[-2,2,1,-1]}\Phi_{02}
    -\kappa\Phi_{22}+\bar{\nu}\Phi_{00}
    +\sigma\Phi_{21}-\bar{\lambda}\Phi_{10}
    -3\delta\Lambda=0\,. \label{eq:Bianchi_div3}
\end{align}
\end{subequations}

Besides the complete set of NP equations above, another set of equations we have used is the linearization of the commutation relations in Eq.~\eqref{eq:commutation_relation}. If one makes the following tetrad choice at $\mathcal{O}(\epsilon)$ in \cite{Campanelli:1998jv, Loutrel:2020wbw}:
\begin{equation} \label{eq:perturbed_tetrad_full}
    l^{\mu(1)}=\frac{1}{2}h_{ll}^{(1)}n^{\mu}\,,\quad
    n^{\mu(1)}=\frac{1}{2}h_{nn}^{(1)}l^{\mu}+h_{ln}^{(1)}n^{\mu}\,,\quad
    m^{\mu(1)}=h_{nm}^{(1)}l^{\mu}+h_{lm}^{(1)}n^{\mu}
    -\frac{1}{2}h_{m\bar{m}}^{(1)}m^{\mu}
    -\frac{1}{2}h_{mm}^{(1)}\bar{m}^{\mu}\,,
\end{equation}
which reduces to Eq.~\eqref{eq:perturbed_tetrad} in the traceful ORG, the linearized commutation relations give us twelve relations between the spin coefficients at  $\mathcal{O}(\epsilon)$ and $h_{ab}^{(1)}$ \cite{Campanelli:1998jv, Loutrel:2020wbw, Wagle:2023fwl}:
\begin{subequations}  \label{eq:perturbed_spin_coefs}
\begin{align}
    & \begin{aligned} \label{eq:reconstruct_kappa}
         \kappa^{(1)}
        =& \;\frac{1}{2}\delta_{[-2,-2,1,1]}h_{ll}^{(1)}
        -D_{[-2,0,0,-1]}h_{lm}^{(1)}\,,
    \end{aligned} \\
    & \begin{aligned} \label{eq:reconstruct_sigma}
        \sigma^{(1)}
        =& \;-\frac{1}{2}D_{[-2,2,1,-1]}h_{mm}^{(1)}
        +(\bar{\pi}+\tau)h_{lm}^{(1)}\,,
    \end{aligned} \\
    & \begin{aligned} \label{eq:reconstruct_lambda}
        \lambda^{(1)}
        =& \;(\pi+\bar{\tau})h_{n\bar{m}}^{(1)}
        +\frac{1}{2}\boldsymbol{\Delta}_{[-1,1,2,-2]}h_{\bar{m}\bar{m}}^{(1)}\,,
    \end{aligned} \\
    & \begin{aligned} \label{eq:reconstruct_nu}
        \nu^{(1)}
        =& \;-\frac{1}{2}\bar{\delta}_{[2,2,-1,-1]}h_{nn}^{(1)}
        +\boldsymbol{\Delta}_{[0,1,2,0]}h_{n\bar{m}}^{(1)}\,,
    \end{aligned} \\
    & \begin{aligned} \label{eq:reconstruct_epsilon}
        \varepsilon^{(1)}
        =& \;\frac{1}{4}\Big[\boldsymbol{\Delta}_{[-1,1,0,-2]}h_{ll}^{(1)}
        -2D_{[0,0,\frac{1}{2},-\frac{1}{2}]}h_{ln}^{(1)}
        -\bar{\delta}_{[-2,0,-3,-2]}h_{lm}^{(1)}
        +\delta_{[-2,0,1,2]}h_{l\bar{m}}^{(1)}
        -(\rho-\bar{\rho})h_{m\bar{m}}^{(1)}\Big]\,,
    \end{aligned} \\
    & \begin{aligned} \label{eq:reconstruct_rho}
        \rho^{(1)}
        =& \;\frac{1}{2}\Big[-\mu h_{ll}^{(1)}
        -(\rho-\bar{\rho})h_{ln}^{(1)}
        -\bar{\delta}_{[-2,0,-1,0]}h_{lm}^{(1)}
        +\delta_{[-2,0,1,2]}h_{l\bar{m}}^{(1)}
        -D_{[0,0,1,-1]}h_{m\bar{m}}^{(1)}\Big]\,,
    \end{aligned} \\
    & \begin{aligned} \label{eq:reconstruct_mu}
        \mu^{(1)}
        =& \;\frac{1}{2}\Big[-\rho h_{nn}^{(1)}
        -\bar{\delta}_{[0,2,-2,-1]}h_{nm}^{(1)}
        +\delta_{[0,2,0,1]}h_{n\bar{m}}^{(1)}
        +(\mu+\bar{\mu})h_{ln}^{(1)}
        +\boldsymbol{\Delta}_{[-1,1,0,0]}h_{m\bar{m}}^{(1)}\Big]\,,
    \end{aligned} \\
    & \begin{aligned} \label{eq:reconstruct_gamma}
        \gamma^{(1)}
        =& \;\frac{1}{4}\Big[-D_{[0,2,1,-1]}h_{nn}^{(1)}
        -\bar{\delta}_{[0,2,-2,-1]}h_{nm}^{(1)}
        +\delta_{[0,2,2,3]}h_{n\bar{m}}^{(1)}
        -(\mu-\bar{\mu}-4\gamma)h_{ln}^{(1)}
        -(\mu-\bar{\mu})h_{m\bar{m}}^{(1)}\Big]\,,
    \end{aligned} \\
    & \begin{aligned} \label{eq:reconstruct_alpha}
        \alpha^{(1)}
        =& \;\frac{1}{4}\Big[-D_{[-2,0,-1,-2]}h_{n\bar{m}}^{(1)}
        +\delta_{[-2,0,1,1]}h_{\bar{m}\bar{m}}^{(1)}
        -\bar{\delta}_{[0,0,-1,-1]}h_{ln}^{(1)}
        +\boldsymbol{\Delta}_{[-2,1,4,-2]}h_{l\bar{m}}^{(1)}
        -\bar{\delta}_{[2,0,-1,-1]}h_{m\bar{m}}^{(1)}\Big]\,,
    \end{aligned} \\
    & \begin{aligned} \label{eq:reconstruct_beta}
        \beta^{(1)}
        =& \;\frac{1}{4}\Big[-D_{[-4,2,2,-1]}h_{nm}^{(1)}
        -\bar{\delta}_{[0,2,-1,-1]}h_{mm}^{(1)}
        -\delta_{[0,0,-1,-1]}h_{ln}^{(1)}
        +\boldsymbol{\Delta}_{[1,2,2,0]}h_{lm}^{(1)}
        +\delta_{[0,-2,1,1]}h_{m\bar{m}}^{(1)}\Big]\,,
    \end{aligned} \\
    & \begin{aligned} \label{eq:reconstruct_pi}
        \pi^{(1)}
        =& \;\frac{1}{2}\Big[D_{[2,0,-1,0]}h_{n\bar{m}}^{(1)}
        +\tau h_{\bar{m}\bar{m}}^{(1)}
        -\bar{\delta}_{[0,0,-1,-1]}h_{ln}^{(1)}
        +\boldsymbol{\Delta}_{[0,1,0,-2]}h_{l\bar{m}}^{(1)}
        +\bar{\tau}h_{m\bar{m}}^{(1)}\Big]\,,
    \end{aligned} \\
    & \begin{aligned} \label{eq:reconstruct_tau}
        \tau^{(1)}
        =& \;\frac{1}{2}\Big[-D_{[0,2,0,-1]}h_{nm}^{(1)}
        +\pi h_{mm}^{(1)}
        +\delta_{[0,0,1,1]}h_{ln}^{(1)}
        -\boldsymbol{\Delta}_{[1,0,-2,0]}h_{lm}^{(1)}
        +\bar{\pi}h_{m\bar{m}}^{(1)}\Big]\,.
    \end{aligned}
\end{align}
\end{subequations}
As mentioned above, we have picked the mostly positive signature, so Eqs.~\eqref{eq:perturbed_tetrad} and \eqref{eq:perturbed_spin_coefs} have an overall opposite sign from the ones in \cite{Loutrel:2020wbw}.
\twocolumngrid

\section{Supplemental Material: Exact relation of the trace subsystem to the corrector-tensor formalism}
\label{sec:ghz_trace_relation}

We show here that Eqs.~\eqref{eq:mu_transport_final} and \eqref{eq:trace_transport} form a first-order factorization of the GHP-prime counterpart of Eq.~(56) of the corrector-tensor formalism in \cite{Green:2019nam}. Let us first translate our equations into the GHP formalism. By definition \cite{Geroch:1973am, Pound:2021qin},
\begin{align} \label{eq:GHP_operators}
    \text{\TH}:=D-p\varepsilon-q\bar{\varepsilon}\,,\;
    \text{\TH}':=\boldsymbol{\Delta}-p\gamma-q\bar{\gamma}\,,\;
    \rho':=-\mu\,,
\end{align}
where $(p,q)$ is the GHP weight of the object on which $\text{\TH}$ and $\text{\TH}'$ act, and the GHP prime operation $f'$ transforms an object $f$ of GHP weight $(p,q)$ into an object of GHP weight $(-p,-q)$. Since $\mu$ has GHP weight $(-1,-1)$, Eqs.~\eqref{eq:mu_transport_final} and \eqref{eq:trace_transport} become
\begin{align}
    \left(\text{\TH}'-2\rho'\right)\rho'^{(1)}
    &=\Phi_{22}^{(1)}\,,
    \label{eq:ghz_mu_ghp} \\
    \left(\text{\TH}'+\rho'-\bar{\rho}'\right)h_{m\bar{m}}^{(1)}
    &=-2\rho'^{(1)}\,.
    \label{eq:ghz_trace_ghp}
\end{align}
Acting on Eq.~\eqref{eq:ghz_trace_ghp} with
$\text{\TH}'-2\rho'$ and using Eq.~\eqref{eq:ghz_mu_ghp} gives
\begin{equation} \label{eq:ghz_composed_operator}
    \left(\text{\TH}'-2\rho'\right)
    \left(\text{\TH}'+\rho'-\bar{\rho}'\right)h_{m\bar{m}}^{(1)}
    =-2\Phi_{22}^{(1)}\,.
\end{equation}
On a Ricci-flat Petrov type-D background, the Ricci identity in Eq.~\eqref{eq:Ricci_Phi22} gives
\begin{equation} \label{eq:ghz_optical_identity}
    \text{\TH}'\rho'=\rho'^2\,,\quad
    \text{\TH}'\bar{\rho}'=\bar{\rho}'^{\,2}\,.
\end{equation}
Consequently, when acting on an object of GHP weight $(0,0)$,
\begin{equation} \label{eq:ghz_operator_identity}
    \left(\text{\TH}'-2\rho'\right)
    \left(\text{\TH}'+\rho'-\bar{\rho}'\right)
    =\text{\TH}'\left(\text{\TH}'-\rho'-\bar{\rho}'\right)
    +2\rho'\bar{\rho}'\,.
\end{equation}
Equation~\eqref{eq:ghz_composed_operator} thus reduces to
\begin{equation} \label{eq:ghz_trace_second_order}
    \left[\text{\TH}'\left(\text{\TH}'-\rho'-\bar{\rho}'\right)
    +2\rho'\bar{\rho}'\right]h_{m\bar{m}}^{(1)}
    =-8\pi T_{nn}^{(1)}\,,
\end{equation}
where we have used that $\Phi_{22}=4\pi T_{nn}$. Accordingly, Eq.~(56) of Ref.~\cite{Green:2019nam} gives\footnote{We have restored the factor of $8\pi G$ absorbed into the definition of $T_{\mu\nu}$ in \cite{Green:2019nam}.}
\begin{equation} \label{eq:ghz_corrector_trace_equation}
    \left[\text{\TH}\left(\text{\TH}-\rho-\bar{\rho}\right)
    +2\rho\bar{\rho}\right]x_{m\bar{m}}^{(1)}
    =8\pi T_{ll}^{(1)}\,,
\end{equation}
where $x_{ab}^{(1)}$ is the corrector tensor in \cite{Green:2019nam}. Since we have used an opposite signature from \cite{Green:2019nam}, our perturbed metric $h_{ab}^{(1)}$ has an opposite sign from the one in \cite{Green:2019nam}. Thus, Eq.~\eqref{eq:ghz_trace_second_order} is precisely the GHP prime of Eq.~\eqref{eq:ghz_corrector_trace_equation}. Thus, our two first-order trace equations [i.e., Eqs.~\eqref{eq:mu_transport_final} and \eqref{eq:trace_transport}] provide a factorization of the second-order GHZ corrector equation. The second-order equation contains two homogeneous functions, corresponding precisely to the integration functions introduced by the two first-order transport equations. Equality of the trace solutions therefore additionally requires these homogeneous data to be matched.

As discussed earlier, Eq.~\eqref{eq:ghz_trace_second_order} or \eqref{eq:ghz_corrector_trace_equation} nevertheless plays different roles in the two formalisms. In the corrector-tensor formalism \cite{Green:2019nam}, $x_{m\bar{m}}^{(1)}$ is the trace component of a
corrector tensor introduced so that the residual source is compatible
with a Hertz-potential reconstruction in the tracefree radiation gauge. In our construction, $h_{m\bar m}^{(1)}$ is the trace component of the
total metric and forms the first branch of a direct reconstruction
hierarchy in a traceful radiation gauge. Thus, this exact overlap
identifies a common trace equation, but not any of the rest of the 
reconstruction algorithm, though the precise relation between these two approaches is worth further investigation.

\twocolumngrid
\section{Supplemental Material: Differential closure of the traceful-ORG hierarchy}

We here show that although the reconstruction hierarchy in Eqs.~\eqref{eq:mu_transport_final}--\eqref{eq:pi_transport}, \eqref{eq:hlm_transport}, \eqref{eq:psi2_transport}, and \eqref{eq:hll_transport} only uses a subset of the Bianchi and Ricci identities in Eqs.~\eqref{eq:Ricci_identities}, \eqref{eq:Bianchi_identities}, and \eqref{eq:Bianchi_div}, the reconstructed metric $h_{ab}^{(1)}$ satisfies the complete linearized Einstein equations. More precisely, we show that given the reconstructed metric $h_{ab}^{(1)}$, the residuals of the linearized Einstein equations in Eq.~\eqref{eq:Einstein_equation} or their trace-reversed form 
\begin{equation} \label{eq:Einstein_equation_reverse}
    \delta R_{ab}[h]
    =8\pi\left(T_{ab}^{(1)}-\frac{1}{2}T^{(1)}g_{ab}^{(0)}\right)\,,\quad
    T^{(1)}\equiv T^{(1)ab}g_{ab}^{(0)}
\end{equation}
satisfy a homogeneous triangular system of transport equations along the principal null direction $n^{\mu}$, where $\delta R_{ab}[h]$ represents the linearized Ricci tensor in $h_{ab}^{(1)}$. Consequently, all of the residuals vanish in a smooth transport domain where the reconstruction hierarchy is nondegenerate, the source is conserved (i.e., $\nabla^{a}T_{ab}^{(1)}=0$), the input $\Psi_4^{(1)}$ satisfies the Teukolsky equation in Eq.~\eqref{eq:Teukolsky_equation}, and the Einstein-residual and Teukolsky-Bianchi compatibility data vanish on a cross-section $\Sigma$ meeting each relevant integral curve of $n^\mu$ once. For the reconstruction route used below, we assume $\mu\neq0$; the case $\mu\equiv0$ and the corresponding adjustment to this proof are treated in the
next section.

For convenience, let us define
\begin{align} \label{eq:residuals}
     & x_{ij}:=\widehat\Phi_{ij}^{(1)}-\Phi_{ij}^{(1)}\,,\;
     L:=\widehat\Lambda^{(1)}-\Lambda^{(1)}\,,\;
     w_k:=\widehat\Psi_k^{(1)}-\Psi_k^{(1)}\,, \nonumber\\
     & i,j\in\{0,1,2\}\,,\quad k\in\{2,3,4\}\,,
\end{align}
where $\widehat{\Phi}_{ij}^{(1)}$ and $\widehat{\Lambda}^{(1)}$ are linearized NP Ricci scalars directly computed from the reconstructed metric $h_{ab}^{(1)}$, while $\Phi_{ij}^{(1)}$ and $\Lambda^{(1)}$ are linearized NP Ricci scalars prescribed by the stress tensor after putting the linearized Einstein equations [i.e., Eq.~\eqref{eq:Einstein_equation_reverse}] on shell. Similarly, $\widehat{\Psi}_k^{(1)}$ are the linearized Weyl scalars computed directly from $h_{ab}^{(1)}$, while $\Psi_k^{(1)}$ are those obtained in the reconstruction hierarchy. No residuals $w_0$ or $w_1$ are introduced since the Weyl scalars $\Psi_{0,1}^{(1)}$ are calculated directly from the reconstructed metric and do not enter the reconstruction hierarchy. Thus, vanishing of all $x_{ij}$ and $L$ is equivalent to the full linearized Einstein equations.

To start, the four Ricci identities in Eqs.~\eqref{eq:Ricci_Phi22}, \eqref{eq:Ricci_Psi4}, \eqref{eq:Ricci_Psi3_Phi21}, and \eqref{eq:Ricci_Psi2_Lambda} used by Eqs.~\eqref{eq:mu_transport}, \eqref{eq:lambda_transport}, \eqref{eq:pi_transport}, and \eqref{eq:Ricci_Psi2_Lambda_linearize} in the reconstruction hierarchy impose that
\begin{equation} \label{eq:selected}
    x_{22}=0\,,\quad w_4=0\,,\quad 
    w_3=-x_{21}\,,\quad w_2=-2L\,.
\end{equation}
When reconstructing $\Psi_3^{(1)}$ via Eq.~\eqref{eq:psi3_transport}, we have used the Bianchi identity Eq.~\eqref{eq:Bianchi_Delta_Psi3}, which implies
\begin{equation}
    \DD_{[4,0,2,0]} w_3-\delta_{[0,4,0,-1]}w_4
    =\DD_{[0,2,2,0]}x_{21}-\bd_{[2,2,0,-1]}x_{22}\,.
\end{equation}
Using Eq.~\eqref{eq:selected}, we obtain
\begin{equation} \label{eq:x21}
    (\DD+2\mu+\bar\mu+2\gamma)x_{21}=0\,.
\end{equation}
Thus, $x_{21}=0$ throughout the transport domain if $x_{21}=0$ on a
cross-section $\Sigma$ transverse to $n^{\mu}$. Furthermore, $x_{12}=\bar{x}_{21}=0$, and Eq.~\eqref{eq:selected} gives $w_3=0$.

Now, subtracting the Teukolsky equation of $\Psi_4^{(1)}$ in Eq.~\eqref{eq:Teukolsky_equation} from the off-shell Teukolsky identity satisfied by the metric-generated $\widehat{\Psi}_4^{(1)}$, and using $w_4=0$ in Eq.~\eqref{eq:selected}, we obtain 
\begin{equation} \label{eq:psi4_closure}
    \cS_4[x]
    =\DD_{[4,1,3,-1]}\mathcal{R}_d[x]
    +\bd_{[3,1,4,-1]}\mathcal{R}_h[x]
    =0\,,
\end{equation}
where $x=\widehat{\Phi}^{(1)}-\Phi^{(1)}$, and $\Phi^{(1)}$ represents all the $\Phi_{ij}^{(1)}$ in $\mathcal{S}_{4}$. Since $x_{22}=x_{21}=0$ if one supplies the correct constraint data on $\Sigma$, the two sources reduce to
\begin{equation}
    \mathcal R_d[x]=-\DD_{[0,1,2,-2]}x_{20}\,,\quad
    \mathcal R_h[x]=0\,.
\end{equation}
Equation~\eqref{eq:psi4_closure} therefore implies
\begin{equation} \label{eq:x20}
    \DD_{[4,1,3,-1]}\DD_{[0,1,2,-2]}x_{20}=0\,.
\end{equation}
This is a homogeneous second-order transport equation.  Hence
$x_{20}=x_{02}=0$ if both homogeneous data vanish. One convenient
formulation is
\begin{equation} \label{eq:x20data}
    x_{20}|_\Sigma=0\,,\quad
    \DD_{[0,1,2,-2]}x_{20}|_\Sigma=0\,.
\end{equation}
Equivalently, the second condition in Eq.~\eqref{eq:x20data} imposes on $\Sigma$ the Bianchi identity Eq.~\eqref{eq:Bianchi_bardelta_Psi3} underlying the derivation of the Teukolsky equation in Eq.~\eqref{eq:Teukolsky_equation}. As Eq.~\eqref{eq:Bianchi_bardelta_Psi3} is not directly used in our reconstruction hierarchy, one can regard this condition as a compatibility requirement on the Teukolsky equation and reconstruction data. Thus, $x_{02}=\bar{x}_{20}=0$.

Finally, when reconstructing $\Psi_2^{(1)}$ from Eq.~\eqref{eq:psi2_transport}, we have used the Bianchi identity Eq.~\eqref{eq:Bianchi_Delta_Psi2}, which implies
\begin{equation} \label{eq:scalarcombination}
    \mu(x_{11}-3L)=0\,,
\end{equation}
where we have used Eq.~\eqref{eq:selected} to replace $w_2$ with $L$ and that $x_{22}=x_{21}=x_{12}=x_{20}=x_{02}=0$. As we have assumed $\mu\ne0$ in the current hierarchy, we obtain
\begin{equation} \label{eq:x11L}
    x_{11}=3L\,.
\end{equation}
Since $\Phi_{11}=(R_{m\bar{m}}+R_{ln})/4$, $\Lambda=(R_{m\bar{m}}-R_{ln})/12$, Eq.~\eqref{eq:x11L} imposes the vanishing of the $ln$ component of the Ricci residual.

In the above discussion, we have shown that our reconstruction hierarchy controls six components of the trace-reversed Einstein equations, i.e.,
\begin{equation} \label{eq:residual_hierarchy}
    x_{22}=x_{21}=x_{12}=x_{20}=x_{02}=x_{11}-3L=0\,,
\end{equation}
which are equivalent to the satisfaction of the following linearized Einstein equations
\begin{equation}
    \{\mathcal{E}_{nn}\,,\mathcal{E}_{nm}\,,
    \mathcal{E}_{n\bar m}\,,\mathcal{E}_{mm}\,,
    \mathcal{E}_{\bar{m}\bar{m}}\,,\mathcal{E}_{m\bar{m}}\}\,,
\end{equation}
where $\mathcal{E}_{ab}$ stands for the $ab$ component of Eq.~\eqref{eq:Einstein_equation}. Notice that, although $x_{11}=3L$ corresponds to the vanishing $ln$ component of the Ricci residual, trace reversal identifies it with $\mathcal{E}_{m\bar{m}}$. The remaining Einstein equations follow from the Ricci-only Bianchi identities in Eq.~\eqref{eq:Bianchi_div}, which reduce to stress-energy conservation once the Ricci tensor is identified with the source.

With Eq.~\eqref{eq:residual_hierarchy}, Eq.~\eqref{eq:Bianchi_div2} reduces to
\begin{equation}
    \DD_{[2,2,0,0]}x_{11}+3\DD L=0\,.
\end{equation}
Using $x_{11}=3L$ gives
\begin{equation} \label{eq:L}
    (\DD+\mu+\bar\mu)L=0\,.
\end{equation}
Thus, $L=x_{11}=w_2=0$ if $L|_\Sigma=0$. Equation~\eqref{eq:Bianchi_div3} then becomes
\begin{equation} \label{eq:x01}
    \DD_{[1,2,-2,0]}x_{01}=0\,,
\end{equation}
so $x_{01}=x_{10}=0$ if $x_{01}|_\Sigma=0$. Finally, Eq.~\eqref{eq:Bianchi_div1} becomes
\begin{equation} \label{eq:x00}
    \DD_{[1,1,-2,-2]}x_{00}=0\,,
\end{equation}
so $x_{00}=0$ if $x_{00}|_\Sigma=0$. Equations~\eqref{eq:L}--\eqref{eq:x00} precisely determine the residuals of the four remaining trace-reversed Einstein equations: $L$, $x_{01}$, $x_{10}$, and $x_{00}$, or equivalently,
\begin{equation}
    \{\mathcal{E}_{ll}\,,\;\mathcal{E}_{lm}\,,\;
    \mathcal{E}_{l\bar{m}}\,,\;\mathcal{E}_{ln}\}\,.
\end{equation}
Furthermore, we have also shown that the Weyl scalars $\Psi_{2,3,4}^{(1)}$ solved from our reconstruction hierarchy agree with the ones directly computed from the reconstructed metric $h_{ab}^{(1)}$. After calculating $\Psi_{0,1}^{(1)}$ from the reconstructed metric, all of the unused Ricci identities then hold as geometric identities. Hence, the Einstein equations in Eq.~\eqref{eq:Einstein_equation} are satisfied throughout the smooth reconstruction domain. For a distributional source, the argument applies separately in each smooth region. Matching, jump, or puncture conditions must additionally ensure that no distributional Einstein residual remains on the source.

\subsection*{Supplemental Material: Rescaled and alternative
reconstruction of $h_{ll}^{(1)}$}
\label{sec:alternative_hll}

We here show how to modify the reconstruction hierarchy above when the background spin coefficient $\mu$ vanishes at a boundary or identically throughout a reconstruction domain.

For the former case, let us write Eq.~\eqref{eq:hll_transport} as
\begin{align}
    & \DD_{[-1,1,-2,-2]}h_{ll}^{(1)}
    =\mathcal{F}^{(1)}+\frac{2}{\mu}\mathcal{S}_\mu^{(1)}\,, \\
    & \mathcal{F}^{(1)}
    \equiv\bar\delta_{[-2,0,-1,0]}h_{lm}^{(1)}
    -\delta_{[-2,0,1,2]}h_{l\bar m}^{(1)} 
    -D_{[0,0,-1,1]}h_{m\bar m}^{(1)}\,,\nonumber
\end{align}
where $\mathcal{S}_\mu^{(1)}$ is the right-hand side of Eq.~\eqref{eq:Ricci_Psi2_Lambda_linearize}. Defining $\tilde{h}_{ll}^{(1)}=\mu h_{ll}^{(1)}$ and using $\DD\mu=-\mu^2-\mu(\gamma+\bar\gamma)$, we obtain
\begin{equation} \label{eq:mu_rescaled_transport}
    \DD_{[0,1,-1,-1]}\tilde{h}_{ll}^{(1)}
    =\mu\mathcal{F}^{(1)}+2\mathcal{S}_\mu^{(1)}\,.
\end{equation}
This is equivalent to Eq.~\eqref{eq:hll_transport} wherever $\mu\neq0$, but contains no explicit division by $\mu$. On a transverse cross-section $\Sigma$ where $\mu\neq0$, the datum $\tilde{h}_{ll}^{(1)}|_\Sigma=(\mu|_\Sigma)h_{ll}^{(1)}|_\Sigma$ replaces the original datum for $h_{ll}^{(1)}$.

Removing the explicit $1/\mu$ factor does not imply that $\tilde{h}_{ll}^{(1)}$ is finite or that the numerical problem is uniformly well conditioned. In the case that $\mu$ vanishes at the Kerr horizon in the Kinnersley tetrad $\left\{l_K^{\mu},n_K^{\mu},m_K^{\mu},\bar{m}_K^{\mu}\right\}$, regularity can instead be assessed in an ingoing chart and a horizon-regular principal tetrad. We use a horizon-regular normalization of the Hawking--Hartle tetrad $\left\{l_H^{\mu},n_H^{\mu},m_H^{\mu},\bar{m}_H^{\mu}\right\}$, with its boost fixed explicitly by
\begin{equation} \label{eq:mu_horizon_boost}
    l_{H}^{\mu}=\mathcal{B}l_{K}^{\mu}\,,\quad
    n_{H}^{\mu}=\mathcal{B}^{-1}n_{K}^{\mu}\,,\quad
    \mathcal{B}=\frac{\Delta}{r^2+a^2}\,,
\end{equation}
and $m_H^{\mu}=m_K^{\mu}$ such that
\begin{equation}
    \mu_{K}
    =-\frac{\Delta}{2\Gamma\bar{\Gamma}^2}\,,\quad
    \mu_{H}
    =\mathcal{B}^{-1}\mu_{K}
    =-\frac{r^2+a^2}{2\Gamma\bar{\Gamma}^2}\,,
\end{equation}
where $\Delta=r^2-2Mr+a^2$ and $\Gamma=r+ia\cos{\theta}$. The coefficient $\mu_{H}$ is finite and nonzero at the future horizon for fixed subextremal spin. The metric components obey
\begin{equation}
    h_{ll}^{H(1)}
    =\mathcal{B}^2h_{ll}^{K(1)}\,,\quad
    h_{l\bar{m}}^{H(1)}
    =\mathcal{B}h_{l\bar{m}}^{K(1)}\,,
\end{equation}
while $h_{\bar{m}\bar{m}}^{(1)}$ and $h_{m\bar{m}}^{(1)}$ are unchanged. One may therefore impose horizon regularity on the components in this regular tetrad, or formulate the reconstruction hierarchy directly in it, transforming all operators, curvature, and source data consistently.

The example involving a Schwarzschild black hole in a shell in the next section illustrates the distinction explicitly. With $f(r)=1-2M/r$, Eqs.~\eqref{eq:spin_coefficients_Schwarzschild} and \eqref{eq:hll_shell} give $\mu=-f/(2r)$ and $h_{ll}^{-(1)}=H(r)/f^2$ in the inner vacuum region adjacent to the horizon, where $H(r)$ is smooth there. Thus, $\tilde{h}_{ll}^{-(1)}=-H/(2rf)$ can diverge at the horizon, where $f = 0$ in Schwarzschild coordinates. Nevertheless, Eq.~\eqref{eq:metric_components_shell}, under the transformation to ingoing Eddington--Finkelstein coordinates $v=t+r_*$ with $dr_*/dr=f^{-1}$, gives $h_{vv}^{-(1)}=H/4$ and $h_{vr}^{-(1)}=h_{rr}^{-(1)}=0$ in that chart. The angular components, determined by Eq.~\eqref{eq:trace_shell}, are also regular at the horizon. The reconstructed metric is therefore regular even though $\tilde{h}_{ll}^{-(1)}$ needs not be finite.

If $\mu\equiv0$ throughout the domain, the rescaled variable $\tilde{h}_{ll}^{(1)}$ cannot determine $h_{ll}^{(1)}$. We instead follow the reconstruction strategy underlying Eq.~(33) of Ref.~\cite{Loutrel:2020wbw}, retaining the trace $h_{m\bar{m}}^{(1)}$ and source terms. Since the background $\mu$ vanishes identically, the left-hand side of Eq.~\eqref{eq:Ricci_Psi2_Lambda_linearize} vanishes, and that identity determines $\Psi_2^{(1)}$ algebraically:
\begin{align} \label{eq:psi2_mu_identically_zero}
    \Psi_2^{(1)}
    =& \;D_{[1,1,0,-1]}\mu^{(1)}
    -\delta_{[-1,1,1,0]}\pi^{(1)}
    -\delta_{[-1,1,1,0]}^{(1)}\pi \nonumber\\
    & \;-2\Lambda^{(1)}\,,
\end{align}
where all the quantities at $\mathcal{O}(\epsilon)$ on the right-hand side have already been reconstructed in the previous steps or are prescribed by the source.

To reconstruct $h_{ll}^{(1)}$, we use the Ricci identity in Eq.~\eqref{eq:Ricci_Phi11_Lambda}. For convenience, let us define
\begin{align} \label{eq:epsilon_hll}
    & E^{(1)}
    \equiv\varepsilon^{(1)}+\bar{\varepsilon}^{(1)}
    =\frac{1}{2}\DD_{[0,0,-1,-1]}h_{ll}^{(1)}
    +\mathcal{W}^{(1)}\,, \nonumber\\
    & \mathcal{J}^{(1)}
    =\mathrm{Re}\left[\Psi_2^{(1)}+\mathcal A^{(1)}\right]
    +\Phi_{11}^{(1)}-\Lambda^{(1)}\,, \nonumber\\
     & \mathcal{W}^{(1)}
    =(\pi+\bar\tau)h_{lm}^{(1)}
    +(\bar\pi+\tau)h_{l\bar{m}}^{(1)}\,, \nonumber\\
    & \begin{aligned}
    \mathcal{A}^{(1)}
    =& \;(\tau+\bar{\pi})\alpha^{(1)}
    +(\bar{\tau}+\pi)\beta^{(1)}
    +\pi\tau^{(1)}+\tau\pi^{(1)} \\
    & \;+\alpha\left(\tau^{(1)}+\bar{\pi}^{(1)}\right)
    +\beta\left(\bar{\tau}^{(1)}+\pi^{(1)}\right)\,,
    \end{aligned}
\end{align}
where the relation for $E^{(1)}$ follows from
Eq.~\eqref{eq:reconstruct_epsilon} and its complex conjugate. The source combinations $\mathcal{J}^{(1)}$, $\mathcal{A}^{(1)}$, and $\mathcal{W}^{(1)}$ contain no unknown $h_{ll}^{(1)}$. Using $\gamma^{(1)}+\bar{\gamma}^{(1)}=0$ and Eq.~\eqref{eq:epsilon_hll}, the real part of the linearized Eq.~\eqref{eq:Ricci_Phi11_Lambda} gives
\begin{equation} \label{eq:hll_division_free}
\begin{aligned}
    \DD_{[0,0,-1,-1]}\DD_{[0,0,-2,-2]}h_{ll}^{(1)}
    =-4\mathcal{J}^{(1)}
    -2\DD_{[0,0,-2,-2]}\mathcal{W}^{(1)}\,,
\end{aligned}
\end{equation}
which is the sourced, traceful generalization of Eq.~(33) in \cite{Loutrel:2020wbw}. The above second-order equation can be turned into two successive first-order transport equations by defining $Z^{(1)}=\DD_{[0,0,-2,-2]}h_{ll}^{(1)}$ such that
\begin{subequations}
\begin{align}
    & \DD_{[0,0,-1,-1]}Z^{(1)}
    =-4\mathcal{J}^{(1)}
    -2\DD_{[0,0,-2,-2]}\mathcal{W}^{(1)}\,, \\
    & \DD_{[0,0,-2,-2]}h_{ll}^{(1)}=Z^{(1)}\,.
\end{align}
\end{subequations}
The initial data $Z^{(1)}|_\Sigma$ and
$h_{ll}^{(1)}|_\Sigma$ on a cross-section $\Sigma$ transverse to the ingoing null direction $n^{\mu}$ are subject to the Einstein and Teukolsky--Bianchi compatibility conditions and the prescribed boundary, source-matching, gauge, and completion conditions.

The derivation of $x_{11}=3L$ in the preceding closure argument must also be modified: when $\mu\equiv0$, Eq.~\eqref{eq:scalarcombination} no longer implies Eq.~\eqref{eq:x11L}. In the residual notation of Eq.~\eqref{eq:residuals}, enforcing Eq.~\eqref{eq:psi2_mu_identically_zero} retains $w_2=-2L$, while Eq.~\eqref{eq:hll_division_free} imposes
\begin{equation}
    \mathrm{Re}[w_2]+x_{11}-L=0\,.
\end{equation}
Since $L$ is real, these relations give $x_{11}=3L$ when $\mu=0$. The preceding arguments for $x_{22}$, $x_{21}$, and $x_{20}$ are unchanged. With $x_{11}=3L$, the remaining residual transport equations in Eqs.~\eqref{eq:L}--\eqref{eq:x00} also apply with $\mu=0$. Thus, source conservation and vanishing residual and Teukolsky--Bianchi compatibility data on $\Sigma$ imply the full linearized Einstein equations throughout the smooth reconstruction domain. For distributional sources, matching, jump, or puncture conditions must additionally remove any residual supported on the source. This provides an alternative reconstruction route of $h_{ll}^{(1)}$ when $\mu$ is identically zero, while preserving the source and compatibility requirements of the original reconstruction hierarchy.

\twocolumngrid
\section{Supplemental Material: Reconstruction of a Schwarzschild black hole in a thin shell}

We here apply the metric reconstruction procedure presented above to reconstruct the background metric correction of a Schwarzschild spacetime when it is surrounded by a thin, static, spherical shell. The \textit{exact} metric for a Schwarzschild black hole surrounded by such a shell was found in coordinates $(t,r,\theta,\phi)$ in \cite{Frauendiener:1990nao, Laeuger:2025zgb}:
\begin{subequations} \label{eq:shell_metric}
\begin{align}
    & ds_+^2=-f_{+}(r)dt^2+f_{+}^{-1}(r)dr^2
    +r^2d\Omega^2\,, \\
    & ds_-^2=-f_{-}(\tilde{r})d\tilde{t}^2
    +f_{-}^{-1}(\tilde{r})d\tilde{r}^2+\tilde{r}^2d\Omega^2\,,
\end{align}
\end{subequations}
where $\tilde{t}$ is linearly proportional to $t$ and $\tilde{r}(r)$ must satisfy
\begin{equation} \label{eq:shell_coordinate}
    \tilde{t}=\frac{\alpha_{+}}{\alpha_{-}}t\,,\quad
    \left(\partial_r\tilde{r}\right)|_{r=R}=\frac{\alpha_{-}}{\alpha_{+}}\,,\quad
    \tilde{r}(r=R)=R\,,
\end{equation}
and $f_{\pm}(r)\equiv1-2M_{\pm}/r$, $M_{-}$ is the black hole's mass $M$, $M_{+}$ is the ADM mass outside the shell, and $\alpha_{\pm}= \sqrt{f_{\pm}(R)}$, with $R$ being the areal radius of the shell. Note that the ADM mass $M_{+}$ outside the shell is not necessarily the black hole's mass $M$ plus the shell's redshifted mass due to the binding energy between the black hole and the shell. When the shell's mass $M_{s}\ll M$, $M_{+}\approx M+\sqrt{f(R)}M_{s}$ \cite{Frauendiener:1990nao}. 

In this work, we work in the perturbative limit $M_{s}\ll M$, so let us define $\delta M\equiv\sqrt{f(R)}M_{s}$ and $\epsilon=\delta M/M$, such that
\begin{equation} \label{eq:mass_linearize}
    M_{+}=M+\epsilon M+\mathcal{O}(\epsilon^2)\,.
\end{equation}
Then, performing the expansion in Eq.~\eqref{eq:metric_expansion} with $g_{\mu\nu}$ being the exact metric of Eq.~\eqref{eq:shell_metric}, we find the background spacetime to be the Schwarzschild metric in Schwarzschild coordinates, with the perturbation
\begin{subequations} \label{eq:background_correction}
\begin{align}
    & \mathfrak{h}^{+(1)}_{tt}
    =f^2(r)\mathfrak{h}^{+(1)}_{rr}
    =\frac{2M}{r}\,,
    \label{eq:background_correction_outside}\\
    & \mathfrak{h}^{-(1)}_{tt}
    =f^2(r)\mathfrak{h}^{-(1)}_{rr}
    =\frac{2M}{Rf(R)}
    \left(1-\frac{3M}{r}+\frac{MR}{r^2}\right)\,,\nonumber\\
    & \mathfrak{h}^{-(1)}_{\phi\phi}
    =\sin^2{\theta}\,\mathfrak{h}^{-(1)}_{\theta\theta}
    =\frac{2M}{Rf(R)}r(r-R)\sin^2{\theta}\,,
    \label{eq:background_correction_inside}
\end{align}
\end{subequations}
where we have chosen $\tilde{r}\approx r+\epsilon M(r-R)/(Rf(R))$ to satisfy Eq.~\eqref{eq:shell_coordinate} and $f(r)\equiv 1-2M/r$. Here and throughout this section, we use $+$ and $-$ to denote the quantities outside and inside the shell, respectively. We also use $\mathfrak{h}_{\mu\nu}^{(1)}$ for the metric perturbation associated with the exact spacetime of \cite{Frauendiener:1990nao}, and $h_{\mu\nu}^{(1)}$ for our reconstructed metric.

To find the stress-energy tensor of the shell, let us apply the second Israel junction condition \cite{Israel:1966rt, Poisson:2009pwt} to the metric inside and outside the shell:
\begin{equation}
    S_{AB}
    =-\frac{1}{8\pi}\left(\left[K_{AB}\right]-[K]q_{AB}\right)\,,
\end{equation}
where $S_{AB}$ is the surface stress-energy tensor of the shell, $K_{AB}\equiv n_{\alpha;\beta}\mathcal{E}^{\alpha}_{A}\mathcal{E}^{\beta}_{B}$ is the extrinsic curvature, $n^{\mu}$ is the normal vector to the shell,
\begin{equation}
    n^{\mu}=(0,\sqrt{f(r)},0,0)\,,
\end{equation}
and $\mathcal{E}^{\alpha}_{A}$ is the pushforward of the intrinsic coordinate $y^{A}=(\tau,\theta,\phi)$ on the shell to the Schwarzschild coordinate $x^{\alpha}=(t,r,\theta,\phi)$, i.e., $\mathcal{E}^{\alpha}_{A}=\partial x^{\alpha}/\partial y^{A}$. We have picked $\tau=\alpha_{+}t=\alpha_{-}\tilde{t}$, which linearizes to $\tau=\sqrt{f(R)}t$ using Eq.~\eqref{eq:mass_linearize}, so the induced metric is $q_{AB}dy^Ady^B=-d\tau^2+R^2d\Omega^2$. Here, we also adopt the notation
\begin{equation}
    \left[Q\right]=Q(R^{+})-Q(R^{-})
\end{equation}
for any function $Q(r)$, where $R^{\pm}$ denote the limits as $r\rightarrow R$ from the exterior and interior side, respectively.

According to \cite{Frauendiener:1990nao}, one finds $S^{AB}$ to take the perfect-fluid form:
\begin{equation}
    S^{AB}=\sigma u^{A}u^{B}+p(q^{AB}+u^{A}u^{B})\,,
\end{equation}
with $u^{A}=(1,0,0)$ being the intrinsic velocity of the shell. The induced energy density $\sigma$ and pressure $p$ of the shell are
\begin{equation}
\begin{aligned}
    & \sigma=\frac{1}{4\pi R}
    \left(\alpha_{-}-\alpha_{+}\right)\,,\\
    & p=\frac{1}{8\pi R^2}
    \left(\frac{R-M_{+}}{\alpha_{+}}
    -\frac{R-M_{-}}{\alpha_{-}}\right)\,,
\end{aligned}
\end{equation}
which after the expansion in Eq.~\eqref{eq:mass_linearize} linearize to
\begin{equation} \label{eq:EOS_linear}
    \sigma=\frac{\epsilon M}{4\pi R^2\sqrt{f(R)}}\,,\quad
    p=\frac{\epsilon M^2}{8\pi R^3 f^{3/2}(R)}\,.
\end{equation}
The full stress-energy tensor $T^{\mu\nu}$ is then given by \cite{Poisson:2009pwt}:
\begin{equation} \label{eq:shell_T}
    T^{\mu\nu}
    =S^{AB}\mathcal{E}^{\mu}_{A}\mathcal{E}^{\nu}_{B}\delta(\ell)
    =\left[\sigma u^\mu u^\nu
    +p\left(\hat{e}_\theta^\mu\hat{e}_\theta^\nu
    +\hat{e}_\phi^\mu\hat{e}_\phi^\nu\right)\right]\delta(\ell)\,,
\end{equation}
where $\ell=(r-R)/\sqrt{f(R)}$ is the proper distance normal to the shell, $\hat{e}^{\mu}_{\theta}$ and $\hat{e}^{\mu}_{\phi}$ are the orthonormal basis vectors in the angular directions, and $u^{\mu}=(f^{-1/2}(r),0,0,0)$ is the four-velocity of the shell.

Our goal is to solve the transport equations in Eqs.~\eqref{eq:mu_transport_final}--\eqref{eq:pi_transport}, \eqref{eq:hlm_transport}, \eqref{eq:psi2_transport}, and \eqref{eq:hll_transport} with the stress-energy tensor in Eq.~\eqref{eq:shell_T} to reconstruct the metric perturbation in Eq.~\eqref{eq:background_correction} in the traceful ORG. On the Schwarzschild background, we use the Kinnersley tetrad:
\begin{equation}
\begin{aligned}
    & l^\mu=\left(\frac{1}{f(r)},1,0,0\right)\,,\quad
    n^\mu=\left(\frac{1}{2},-\frac{f(r)}{2},0,0\right)\,, \\
    & m^\mu=\frac{1}{\sqrt{2}r}
    \left(0,0,1,i\csc{\theta}\right)\,.
\end{aligned}   
\end{equation}
In this case, the only nonzero spin coefficients are \cite{Chandrasekhar_1983}
\begin{equation} \label{eq:spin_coefficients_Schwarzschild}
    \rho=-\frac{1}{r}\,,\;
    \mu=-\frac{f(r)}{2r}\,,\;
    \gamma=\frac{M}{2r^2}\,,\;
    \alpha=-\beta=-\frac{\cot{\theta}}{2\sqrt{2}r}\,,
\end{equation}
and $\Psi_2=-Mr^{-3}$ is the only nonzero Weyl scalar. Thus, the only nonzero components of $T_{\mu\nu}^{(1)}$ in the NP basis are:
\begin{equation}
\begin{aligned}
    & T_{ll}^{(1)}=\frac{M}{4\pi R^2f(R)}\delta(r-R)\,,
    && T_{nn}^{(1)}=\frac{M f(R)}{16\pi R^2}\delta(r-R)\,, \\
    & T_{ln}^{(1)}=\frac{M}{8\pi R^2}\delta(r-R)\,,
    && T_{m\bar m}^{(1)}=\frac{M^2}{8\pi R^3f(R)}\delta(r-R)\,.
\end{aligned}    
\end{equation}
Since $\Phi_{00}=R_{ll}/2$, $\Phi_{22}=R_{nn}/2$, $\Phi_{11}=(R_{m\bar{m}}+R_{ln})/4$, $\Lambda=(R_{m\bar{m}}-R_{ln})/12$, and $R_{\mu\nu}=8\pi(T_{\mu\nu}-g_{\mu\nu}T/2)$, the only nonzero NP Ricci scalars at $\mathcal{O}(\epsilon)$ are
\begin{equation} \label{eq:Phi_shell}
\begin{aligned}
    & \Phi_{00}^{(1)}
    =\frac{M}{R^2f(R)}\delta(r-R)\,,\\
    & \Phi_{22}^{(1)}
    =\frac{M f(R)}{4R^2}\delta(r-R)\,,\\
    & \Phi_{11}^{(1)}
    =\frac{M(R-M)}{4R^3f(R)}\delta(r-R)\,,\\
    & \Lambda^{(1)}
    =\frac{M(R-3M)}{12R^3f(R)}\delta(r-R)\,.
\end{aligned}
\end{equation}

Let us first examine the spin-2 (i.e., $h_{\bar{m}\bar{m}}^{(1)}$) and spin-1 (i.e., $h_{l\bar{m}}^{(1)}$) parts of the background metric correction. Since $T_{nm}^{(1)}=T_{mm}^{(1)}=0$ and $T_{nn}^{(1)}$ is purely radial, one can easily verify from \cite{Teukolsky:1973ha} that the source in the stationary Teukolsky equation of $\Psi_4^{(1)}$ vanishes. Since $\Psi_4^{(1)}$ is a spin-2 field, and we are interested in spherically symmetric corrections to the background spacetime due to the shell, we need to set the homogeneous solution $\Psi_4^{(1)}=0$. As Eqs.~\eqref{eq:lambda_transport} and \eqref{eq:hmm_transport} also impose $\lambda^{(1)}$ and $h_{\bar{m}\bar{m}}^{(1)}$ to be spin-2 fields, $\lambda^{(1)}=h_{\bar{m}\bar{m}}^{(1)}=0$ for the same reason. Similarly, since $\Phi_{21}^{(1)}=0$ and $\Phi_{22}^{(1)}$ is purely radial, one finds from Eq.~\eqref{eq:Bianchi_Delta_Psi3_source} that $\mathcal{R}_{h}^{(1)}=0$. Thus, Eq.~\eqref{eq:psi3_transport} and spherical symmetry impose $\Psi_{3}^{(1)}=0$. Furthermore, since $\pi=\tau=0$ in a Schwarzschild spacetime, the source of Eq.~\eqref{eq:pi_transport} also vanishes, so we set $\pi^{(1)}=0$, which leads to $h_{l\bar{m}}^{(1)}=0$ via Eq.~\eqref{eq:hlm_transport}.

The only nonzero components of $h_{ab}^{(1)}$ are then spin-0 contributions: $h_{m\bar{m}}^{(1)}$ and $h_{ll}^{(1)}$. For $h_{m\bar{m}}^{(1)}$, Eqs.~\eqref{eq:mu_transport_final} and \eqref{eq:trace_transport} reduce to
\begin{subequations}
\begin{align}
    & \left(\frac{f(r)}{2}\partial_r+\frac{r-3M}{r^2}\right)\mu^{(1)}
    =\Phi_{22}^{(1)}\,, \\
    & \frac{f(r)}{2}\partial_r h_{m\bar{m}}^{(1)}=-2\mu^{(1)}\,,
    \label{eq:trace_transport_shell}
\end{align}
\end{subequations}
which can be decoupled into
\begin{equation} \label{eq:hmm_eq_shell}
    \partial_r\left(r^2\partial_r h_{m\bar{m}}^{(1)}\right)
    =-\frac{8r^2}{f^2(r)}\Phi_{22}^{(1)}
    =-\frac{2M}{f(R)}\delta(r-R)\,.
\end{equation}
The most general solution of Eq.~\eqref{eq:hmm_eq_shell} is
\begin{equation}
    h_{m\bar{m}}^{+}=c_0^{+}-\frac{c_1^{+}}{r}\,,\quad
    h_{m\bar{m}}^{-}=c_0^{-}-\frac{c_1^{-}}{r}\,,
\end{equation}
where $c_{0,1}^{\pm}$ are constants. Let us impose $c_0^{+}=c_1^{+}=0$, which is equivalent to fixing the outside spacetime to still be in Schwarzschild form. Then, Eq.~\eqref{eq:hmm_eq_shell} gives the jump condition
\begin{equation}
    \left[r^2\partial_rh_{m\bar{m}}^{(1)}\right]
    =-\frac{2M}{f(R)}\,,
\end{equation}
which, along with the continuity condition for $h_{m\bar{m}}^{(1)}$, implies that
\begin{equation} \label{eq:trace_shell}
    h_{m\bar{m}}^{(1)}
    =\frac{2M}{Rf(R)}\left(1-\frac{R}{r}\right)\Theta(R-r)\,.
\end{equation}

To find $h_{ll}^{(1)}$, let us first solve the transport equation of $\Psi_2^{(1)}$ in Eq.~\eqref{eq:psi2_transport}, which in Schwarzschild spacetime reduces to 
\begin{equation} \label{eq:psi2_eq_shell}
    \partial_r(r^3\Psi_2^{(1)})
    =-\frac{2r^3}{f(r)}\mathcal{R}_{g}^{(1)}-\frac{6M}{f(r)}\mu^{(1)}\,,
\end{equation}
where
\begin{equation}
    \mathcal{R}_g^{(1)}
    =-\frac{M(2R-3M)}{12R^3}\delta'(r-R)
    +\frac{M^2}{12R^4 f(R)}\delta(r-R)\,.
\end{equation}
From Eqs.~\eqref{eq:trace_transport_shell} and \eqref{eq:trace_shell}, we also obtain that
\begin{equation} \label{eq:mu_shell}
    \mu^{(1)}=-\frac{Mf(r)}{2f(R)r^2}\Theta(R-r)\,.
\end{equation}
Integrating Eq.~\eqref{eq:psi2_eq_shell}, we obtain
\begin{equation} \label{eq:psi2_shell}
\begin{aligned}
   \Psi_2^{(1)}
    =& \;\frac{3M^2}{Rf(R)}
    \frac{r-R}{r^4}\Theta(R-r)
    -\frac{M}{r^3}\Theta(r-R) \\
    & \;+\frac{M(2R-3M)}{6R^3 f(R)}\delta(r-R)\,, 
\end{aligned}  
\end{equation}
where we have used the distribution identity:
\begin{equation}
    f(r)\delta'(r-R)=f(R)\delta'(r-R)-f'(R)\delta(r-R)\,,
\end{equation}
and imposed $\Psi_{2,\mathrm{reg}}^{+(1)}=-M/r^{3}$ and $[\Psi_{2,\mathrm{reg}}^{(1)}]=-M/R^{3}$ to fix the integration constant, where $\Psi_{2,\mathrm{reg}}^{(1)}$ denotes the non-distributional part of $\Psi_2^{(1)}$. This is essentially imposing that the spacetime be Schwarzschild with a mass shift across the shell.

Now, let us examine Eq.~\eqref{eq:hll_transport}, which in a Schwarzschild background spacetime reduces to
\begin{equation} \label{eq:hll_eq_shell_1}
\begin{aligned}
    & \left(\frac{f(r)}{2}\partial_r+\frac{2M}{r^2}\right)h_{ll}^{(1)} \\
    & =\partial_r h_{m\bar{m}}^{(1)}
    +\frac{2}{\mu}\left[\left(\partial_r+\frac{1}{r}\right)\mu^{(1)}
    -\Psi_2^{(1)}-2\Lambda^{(1)}\right] \\
    & =-\frac{8}{f(r)}\mu^{(1)}+
    \frac{2}{\mu}\left(\partial_r\mu^{(1)}-\Psi_2^{(1)}-2\Lambda^{(1)}\right)\,,
\end{aligned}
\end{equation}
where we have used Eq.~\eqref{eq:trace_transport_shell} to find the last line. From Eqs.~\eqref{eq:Phi_shell}, \eqref{eq:mu_shell}, and \eqref{eq:psi2_shell}, one finds that
\begin{equation}
\begin{aligned}
    & \partial_r\mu^{(1)}-\Psi_2^{(1)}-2\Lambda^{(1)} \\
    & =\frac{M(R-3M)}{f(R)Rr^3}\Theta(R-r)
    +\frac{M}{r^3}\Theta(r-R)\,,
\end{aligned}
\end{equation}
where the $\delta(r-R)$ terms all cancel exactly. Then, Eq.~\eqref{eq:hll_eq_shell_1} reduces to
\begin{equation} \label{eq:hll_eq_shell_2}
\begin{aligned}
    & \left(\frac{f(r)}{2}\partial_r+\frac{2M}{r^2}\right)h_{ll}^{(1)} \\
    & =-\frac{4M}{r^2f(r)}\Theta(r-R)
    +\frac{4M^2(3r-2R)}{f(R)f(r)R r^3}\Theta(R-r)\,.
\end{aligned}
\end{equation}
Solving Eq.~\eqref{eq:hll_eq_shell_2} gives the particular solutions displayed below together with homogeneous terms $C_\pm/f^2(r)$. Continuity across the shell imposes $C_+=C_-\equiv C$, while the exterior asymptotic time normalization, equivalently the absence of a nondecaying contribution to $h_{tt}^{+(1)}$ at infinity, fixes $C=0$. One then gets
\begin{equation} \label{eq:hll_shell}
\begin{aligned}
    & h_{ll}^{+(1)}=\frac{8M}{rf^2(r)}\,, \\
    & h_{ll}^{-(1)}=\frac{8M}{Rf^2(r)f(R)}
    \left(1-\frac{3M}{r}+\frac{MR}{r^2}\right)\,.
\end{aligned}
\end{equation}
At $r=R$, one can verify that $h_{ll}^{+(1)}(R)=h_{ll}^{-(1)}(R)=8M/(Rf^2(R))$, so $h_{ll}^{(1)}$ is continuous and free of distributional singularities.

Now, let us compare our results in Eqs.~\eqref{eq:trace_shell} and \eqref{eq:hll_shell} via metric reconstruction to the ones in Eq.~\eqref{eq:background_correction}. Since we have chosen the traceful ORG in Eq.~\eqref{eq:radiation_gauges_traceful} for the reconstruction, we need to find the proper coordinate transformation from our results to Eq.~\eqref{eq:background_correction}. First, the traceful ORG condition $h_{\mu\nu}^{(1)}n^\nu=0$ implies that
\begin{equation}
    h_{tt}^{(1)}=f(r)h_{tr}^{(1)}
    =f^2(r)h_{rr}^{(1)}\,,
\end{equation}
so
\begin{equation}
    h_{ll}^{(1)}
    =f^{-2}(r)h_{tt}^{(1)}+2f^{-1}(r)h_{tr}^{(1)}+h_{rr}^{(1)}
    =4h_{rr}^{(1)}\,.
\end{equation}
Furthermore,
\begin{equation}
    h_{m\bar{m}}^{(1)}
    =\frac{1}{2r^2}(h_{\theta\theta}^{(1)}
    +\csc^2\theta h_{\phi\phi}^{(1)})\,,
\end{equation}
so after imposing the spherical symmetry condition $h_{\theta\phi}^{(1)}=0$ and $h_{\phi\phi}^{(1)}=\sin^2{\theta}h_{\theta\theta}^{(1)}$, we find
\begin{equation} \label{eq:metric_components_shell}
\begin{aligned}
    & h_{tt}^{(1)}=f(r)h_{tr}^{(1)}
    =f^2(r)h_{rr}^{(1)}=\frac{f^{2}(r)}{4}h_{ll}^{(1)}\,, \\
    & h_{\phi\phi}^{(1)}=\sin^2{\theta}\,h_{\theta\theta}^{(1)}
    =r^2\sin^2\theta h_{m\bar{m}}^{(1)}\,.
\end{aligned}
\end{equation}
Using the solutions in Eqs.~\eqref{eq:trace_shell} and \eqref{eq:hll_shell}, we find the nonzero metric perturbation components outside the shell to be
\begin{equation} \label{eq:h_shell_outside}
    h_{tt}^{+(1)}=f(r)h_{tr}^{+(1)}
    =f^2(r)h_{rr}^{+(1)}=\frac{2M}{r}\,,\\  
\end{equation}
while the ones inside the shell are
\begin{equation} \label{eq:h_shell_inside}
\begin{aligned}
    h_{tt}^{-(1)}
    =& \;f(r)h_{tr}^{-(1)}=f^2(r)h_{rr}^{-(1)} \\
    =& \;\frac{2M}{Rf(R)}
    \left(1-\frac{3M}{r}+\frac{MR}{r^2}\right)\,, \\
    h_{\phi\phi}^{-(1)}
    =& \;\sin^2{\theta}\,h_{\theta\theta}^{-(1)}
    =\frac{2M}{Rf(R)}r(r-R)\sin^2\theta\,.
\end{aligned}    
\end{equation}

Now, let us consider separate coordinate transformations $x^{\mu}_{\pm}\rightarrow x^{\mu}_{\pm}+\epsilon\xi^{\mu(1)}_{\pm}$ outside and inside the shell such that $h_{\mu\nu}^{\pm(1)}\rightarrow h_{\mu\nu}^{\pm(1)}-2\xi_{(\mu;\nu)}^{\pm(1)}$. Since we want to avoid introducing any components in $\{h_{t\theta}^{\pm(1)}$, $h_{t\phi}^{\pm(1)},h_{r\theta}^{\pm(1)}$, $h_{r\phi}^{\pm(1)},h_{\theta\phi}^{\pm(1)}\}$ and keep $h_{\theta\theta}^{\pm(1)}$ free of angular dependence after the transformation, we set $\xi_{\theta}^{\pm(1)}=\xi_{\phi}^{\pm(1)}=0$ while $\xi_{t}^{\pm(1)}$ and $\xi_{r}^{\pm(1)}$ to be purely radial. Defining $\delta_\xi h_{\mu\nu}^{(1)}=-2\xi_{(\mu;\nu)}^{(1)}$, one then finds that
\begin{equation} \label{eq:h_gauge_transform}
\begin{aligned}
    & \delta_\xi h_{tt}^{(1)}
    =f(r)f'(r)\,\xi_r^{(1)}\,,\;
    \delta_\xi h_{rr}^{(1)}=
    -2\left(\partial_r
    +\frac{f'(r)}{2f(r)}\right)\xi_r^{(1)}\,, \\
    & \delta_\xi h_{tr}^{(1)}
    =-\left(\partial_r-\frac{f'(r)}{f(r)}\right)\xi_t^{(1)}\,,\\
    & \delta_\xi h_{\phi\phi}^{(1)}
    =\sin^2{\theta}\delta_\xi h_{\theta\theta}^{(1)}
    =-2r\sin^2{\theta}f(r)\xi_r^{(1)}\,.
\end{aligned}    
\end{equation}

Outside the shell, comparing Eq.~\eqref{eq:h_shell_outside} to Eq.~\eqref{eq:background_correction_outside}, we find $h_{tt}^{+(1)}=\mathfrak{h}_{tt}^{+(1)}$ and $h_{rr}^{+(1)}=\mathfrak{h}_{rr}^{+(1)}$, so we only need to remove $h_{tr}^{+(1)}$. From Eq.~\eqref{eq:h_gauge_transform}, we can set
\begin{equation} \label{eq:xi_r_+}
\begin{aligned}
    \xi_{r}^{+(1)}=0\,,\quad
    \left(\partial_r-\frac{f'(r)}{f(r)}\right)\xi_t^{+(1)}
    =\frac{2M}{rf(r)}\,.
\end{aligned}
\end{equation}
The most general solution to the differential equation of $\xi_{t}^{+(1)}$ above is
\begin{equation} \label{eq:xi_t_+}
    \xi_t^{+(1)}
    =f(r)\left[C_{t}^{+}+2M
    \log{\left(\frac{r}{2M}-1\right)}\right]
    -\frac{4M^2}{r}\,,
\end{equation}
where $C_t^{+}$ is an integration constant.

Similarly, inside the shell, all the components in Eq.~\eqref{eq:h_shell_inside} are consistent with the ones in Eq.~\eqref{eq:background_correction_inside} except $h_{tr}^{(-)}\neq0$, so we remove $h_{tr}^{(-)}$ by setting
\begin{equation} \label{eq:xi_r_-}
\begin{aligned}
    & \xi_{r}^{-(1)}=0\,,\\
    & \left(\partial_r-\frac{f'(r)}{f(r)}\right)\xi_t^{-(1)}
    =\frac{2M}{Rf(R)f(r)}
    \left(1-\frac{3M}{r}+\frac{MR}{r^2}\right)\,.
\end{aligned}
\end{equation}
The most general solution to the differential equation of $\xi_t^{-(1)}$ above is
\begin{equation} \label{eq:xi_t_-}
\begin{aligned}
    \xi_t^{-(1)}
    =& \;f(r)\left[C_{t}^{-}
    +\frac{2M}{Rf(R)}
    \left(r+M\log{\left(\frac{r}{2M}-1\right)}\right)\right] \\
    & \;-\frac{2M^2}{r}\,,
\end{aligned}
\end{equation}
where $C_t^{-}$ is an integration constant. Thus, by performing the gauge transformations in Eqs.~\eqref{eq:xi_r_+}--\eqref{eq:xi_t_-}, one finds $h_{\mu\nu}^{\pm(1)}$ in Eqs.~\eqref{eq:h_shell_outside} and \eqref{eq:h_shell_inside} to be exactly the same as $\mathfrak{h}_{\mu\nu}^{\pm(1)}$ in Eqs.~\eqref{eq:background_correction_outside} and \eqref{eq:background_correction_inside}.

\bibliographystyle{apsrev4-2}
\bibliography{reference}

\end{document}